\documentstyle[aps,prb,multicol,epsf]{revtex}

\let\mbf\bf   

\def\Avec{{\mbf A}}
\def\evec{{\mbf e}}
\def\kvec{{\mbf k}}
\def\qvec{{\mbf q}}
\def\pvec{{\mbf p}}

\def\Re{\mathop{\rm Re}\nolimits}
\def\Im{\mathop{\rm Im}\nolimits}
\def\abs#1{{\left|#1\right|}}
\def\absii#1{{\left|#1\right|}^2}
\def\expv#1{{\left\langle#1\right\rangle}}
\def\braket#1#2#3{\left\langle#1\mid#2\mid#3\right\rangle}

\def\eV{{\rm eV}} \def\meV{{\rm meV}}
\def\waven{{\rm cm}^{-1}}

\def\chiScr{{\chi_{\rm Scr}}}
\def\Relam{{\lambda'}}
\def\Imlam{{\lambda''}}
\def\dxiiyii{{d_{x^2-y^2}}}

\def\undergl#1#2{\raise1pt\vtop{\baselineskip0pt\lineskip0pt
  \ialign{$\mathsurround0pt#1\hfil##\hfil$\crcr#2\crcr\sim\crcr}}}
\def\appgeq{\mathrel{\mathpalette\undergl>}}

\begin{document}


\title {Electronic Raman scattering in YBCO\\
and other superconducting cuprates}

\author {T. Strohm and M. Cardona}

\address {Max-Planck-Institut f\"ur Festk\"orperforschung, 
Heisenbergstr.~1, D-70569 Stuttgart, Germany}

\date{\today}

\maketitle


\begin{abstract}

Superconductivity induced structures in the electronic Raman spectra
of high-$T_c$ superconductors are computed using the results of {\it
ab initio} LDA-LMTO three-dimensional band structure calculations via
numerical integrations of the mass fluctuations, either in the whole 3D
Brillouin zone or limiting the integrations to the Fermi surface. The
results of both calculations are rather similar, the Brillouin zone
integration yielding additional weak structures related to the
extended van Hove singularities. Similar calculations have been
performed for the normal state of these high-$T_c$ cuprates.
Polarization configurations have been investigated and the results
have been compared to experimental spectra. The assumption of a simple
$\dxiiyii$-like gap function allows us to explain a number of
experimental features but is hard to reconcile with the relative
positions of the $A_{1g}$ and $B_{1g}$ peaks.

\end{abstract}



\begin{multicols}{2}

\section{Introduction}
\label{sec:intro}

 
After ten years of research in the field of high-$T_c$
superconductors\cite{BedMue} (HTSC), many of their properties have not
yet been understood.  In particular, the symmetry of the
superconducting gap\cite{Lyo,Dyn,Sch} is still controversial. Usually,
one assumes that the superconducting condensate in the HTSC can be
described by an order parameter $\Delta_\kvec$, which depends only on
the quasi momentum $\kvec$, but not on band index $n$.  Retardation
effects are also often neglected, i.e.\ the gap is assumed to be
independent of the frequency $\omega$.

A wide range of experimental techniques can be employed to investigate
the properties of the gap function. Among these, Raman scattering has
played an important role.\cite{DevEin,KraCar} The dependence of the
Raman response on the directions of polarization of the incident and
scattered light yields several independent spectra which provide a
considerable number of constraints on the assumed $\kvec$-dependence
of the gap function $\Delta_\kvec$.  However, Raman scattering is not
sensitive to the phase of the gap.

The Raman spectra at temperatures below $T_c$ shows, in most HTSC, a
clear gap-like structure which lies in the energy range of the optical
phonons at the ${\it \Gamma}$ point. These phonons have been
identified for most HTSC,\cite{ThoCar} and the subtraction of the
corresponding structures from the spectra has become a standard
procedure to isolate electronic structures containing gap
information. Electronic Raman scattering spectra are now available for
many high-$T_c$ materials and, since they exhibit similar general
features, most of these data are considered to be reliable. In this
paper, we attempt to interpret these spectra from a theoretical point
of view based on the full 3D one-electron band structure. We pay
attention to both, line shapes and {\it absolute} scattering
efficiencies.

The theory of electronic Raman scattering in superconductors was
pioneered by Abrikosov and coworkers in two important
papers.\cite{AbrFal,AbrGen} In the first, they developed a theory for
the scattering efficiency of {\it isotropic} Fermi liquids under the
assumption that the attractive interaction between quasiparticles can
be neglected. In the second paper, they extended this approach to
anisotropic systems, introduced the effective mass vertex concept, and
included Coulomb screening.  The current form of the theory, developed
mainly by Klein {\it et al.},\cite{KleDie} takes into account the
attractive pairing interaction and emphasizes the role of gauge
invariance as well as the polarization dependence for anisotropic
gaps.  In order to compare the theoretical predictions with the
experiment, we evaluate them numerically in a quantitative manner
(including {\it absolute} scattering efficiencies!) and compare them
to the experimental findings.

Several calculations of the electronic Raman scattering efficiency of
HTSC have already been published. Some of them use highly simplified
2D band structures and a decomposition of the Raman vertex
$\gamma_\kvec$ in Fermi surface (FS) harmonics\cite{Allen} or
Brillouin zone (BZ) harmonics, as well as FS integrations instead of
the required BZ integrations.\cite{DevEin,DevRepl,DevEinPRB} The
results of these calculations depend very strongly on the number of
expansion coefficients used for $\gamma_\kvec$ and their relative
values.  Another approach\cite{KraMaz} involves the use of band
structures calculated in the framework of the local density
approximation\cite{KohSha} (LDA) using the LMTO
method.\cite{AndJep,AndLie} Within the approximations of the LDA, this
Raman vertex is exact, i.e.\ the only errors made in such a
calculation arise from limitations of the LDA method itself and from
the discretization of the Brillouin zone or Fermi surface. Some of
these calculations, however, suffer from the fact that only the
imaginary part of the Tsuneto function\cite{Tsu} has been used, and
that only 2D integrations were performed.\cite{KraCar}


The present approach\cite{CarStr} is based on the full 3D LDA-LMTO
band structure. It uses a BZ integration, screening effects are
included, and both the real and imaginary part of the Tsuneto function
are used as required by the theory. Electronic Raman spectra are
calculated for ${\rm YBa}_2{\rm Cu}_3{\rm O}_7$ (Y-123) and ${\rm
YBa}_2{\rm Cu}_4{\rm O}_8$ (Y-124). The orthorhombicity of the
cuprates is also taken into account in the Raman vertex since we use
as starting point the band structure of the {\it orthorhombic}
materials.


The cuprates under consideration are not only of interest because of
their superconducting, but also of their strange normal-conducting
properties. Usual metals should show peaks in their Raman spectra at
their plasma frequencies corresponding to Raman shifts of a few
$\eV$. The optimally doped cuprates, in contrast, show a very broad
electronic background (from $0$ to about $1\,\eV$ Raman shift), which
is almost independent of temperature and frequency. The spectra of the
underdoped HTSC, such as Y-124, show some temperature dependence at
low frequencies ($\hbar\omega\ll kT$). It is possible to explain these
peculiarities, together with other properties, by assuming a certain
form of the quasiparticle lifetime, as was done in the Marginal Fermi
Liquid theory.\cite{Var,VarII}


For the superconducting state, various forms for the gap function have
been proposed. That which has received most experimental support has
$\dxiiyii$ symmetry, i.e., $B_{1g}$ symmetry in tetragonal HTSC. The
power of Raman scattering to confirm such gap function has been
questioned, because, among other difficulties to be discussed below,
it only probes the {\it absolute} value of the gap function, i.e.\ it
cannot distinguish between a $\dxiiyii$-like gap function (for
instance $\cos2\phi$), and a $\left|\cos2\phi\right|$ gap function,
which corresponds to anisotropic $s$ ($A_{1g}$) symmetry.  However, it
was pointed out that addition of impurities can be used to effect
the distinction.\cite{DevImp}

This paper is organized as follows: in Sec.~\ref{sec:ldabands}, we
review the main properties of the band structures of the investigated
cuprates, as obtained by LDA-LMTO calculations.
Sec.~\ref{sec:gtheory} discusses the theory of electronic Raman
scattering in systems with an anisotropic band structure. We first
introduce the basic concept of Raman vertex, and then present an
expression relating the scattering efficiency to the Raman
susceptibility. In Sec.~\ref{sec:stheory} and~\ref{sec:ntheory} we
derive expressions for the Raman susceptibility in the superconducting
and normal conducting phases, and discuss some effects not directly
contained in the presented form of the theory. Section~\ref{sec:expts}
is concerned with aspects of the experiments which have to be taken
care of, especially with regard to the comparison with the
theory. Finally, in Sec.~\ref{sec:numerics} the results of our
numerical calculation are presented and compared to the experimental
results. The difference between calculations involving only FS
averaging, and those in which such averaging is performed over the
whole BZ, are discussed.

\section{LDA band structure}
\label{sec:ldabands}

The basis of our calculation is the LDA-LMTO band structure of the
HTSC under consideration.\cite{AndLie} For the sake of further
discussion, we shall describe briefly such band structure.

The Fermi surface of $\rm YBa_2Cu_3O_7$ (Y-123)\cite{AndLie} consists
of four sheets, an even and an odd $pd\sigma$-like plane band, a
$pd\sigma$-like chain sheet and a very small $pd\pi$-like chain
sheet. The latter is predicted by the full-potential LMTO calculations
as well as LAPW calculations.\cite{OKAnd} We use the atomic spheres
approximation (ASA) to the LMTO, which does not reproduce this rather
small feature.  In the case of Y-\nobreak123 the three $pd\sigma$-like
conduction bands extend from $-1\,\eV$ to $2\,\eV$ relative to the
Fermi energy. They are embedded in a broad valence band, which ranges
from $-7\,\eV$ to $2\,\eV$ and consists of 36 bands (mainly Cu-$d$ and
O-$p$ orbitals). Below $-7\,\eV$, there is a gap of $4\,\eV$. Above
the conduction band, there is another gap of $0.5\,\eV$, above which
are the lowest fully unoccupied bands which consist mainly of $d$
orbitals of Y and Ba.

The band structure of $\rm YBa_2Cu_4O_8$ (Y-124) shows similar
features. There is an additional $pd\sigma$-like chain band, while the
$pd\pi$-like chain bands are predicted by both, full-potential LMTO and
LAPW to contain no holes, i.e.\ to be completely filled.

An interesting feature of the band structure of both Y-123 and Y-124
is an extended saddle point\cite{AndJep} on the $k_x$-axis near the
$X$ point. This extended saddle point corresponds to a van Hove
singularity at approximately $25\,\meV$ (Y-123) and $110\,\meV$
(Y-124), respectively, below the Fermi level. As will be shown, the
comparatively large density of states in this energy region and the
warped nature of the corresponding bands has an influence on the
calculated electronic Raman spectrum.

The band structure which we used in our numerical calculations was
evaluated for Y-123 on a mesh of $48\times48\times12$ points in the
first BZ, involving $4373$ irreducible points. The band structure of
Y-124 is less sensitive to the resolution of the grid (because the
extended saddle point lies deeper with respect to the Fermi
surface). It was thus sufficient to use a $24\times24\times12$ mesh
with $1099$ irreducible points. The calculations of the
self-consistent potential have been performed in the ASA. Therefore,
the $pd\pi$ chain band around the $S$ point, which should be partly
filled, is completely filled. Because of the small number of states
involved, we do not think this should affect significantly our
results.

As stressed above, our calculations are based on a band structure
obtained within the LDA. We are aware of the fact, that the mean free
path for transport in the direction of the $c$-axis is smaller than
the size of the unit cell, i.e.\ that a description by means of a band
structure $\epsilon_{n\kvec}$ may be questionable
(Ioffe-Regel-limit). Nevertheless a nontrivial band structure in the
$c$-direction may simulate some of the $c$-direction confinement
effects and represent, after integration along $k_z$, a reasonable 2D
band structure.

\section{General theory}
\label{sec:gtheory}

Two approaches have been used to derive the cross section (called
scattering efficiency when referred to unit path length in the solid)
of electronic Raman scattering in superconductors with anisotropic
Fermi surfaces. The first uses Green's
functions,\cite{AbrFal,AbrGen,KleDie,Kos} and the second the kinetic
equation.\cite{Woe,DevEin} Both start with the simplification of the
Hamiltonian, using $\kvec\cdot\pvec$ theory, which relates the Raman
vertex $\gamma_\kvec$ to the inverse effective mass
tensor.\cite{AbrGen} We first briefly review this procedure and,
subsequently, the derivation of the expression for the scattering
efficiency using the diagrammatic approach.


\subsection{The Raman vertex}

To derive an expression for the efficiency for electronic Raman
scattering, one has to replace the momentum $\pvec$ in the Hamiltonian
by $\pvec-(e/c)\Avec$.  This yields two distinct perturbation terms:
$H_{AA}=(r_0/2)\Avec^2$, quadratic in the vector potential, and
$H_A=-(e/mc)\Avec\cdot\pvec$, linear in $A$ (we use the transverse
gauge; $r_0=e^2/(m_0c^2)$ denotes the classical electron radius). The
relevant states in the theory are composed of the state of electrons
in the sample plus the state of the photon field. The initial state of
the photon field has $n_L$ laser photons with wave vector $\kvec_L$
and polarization $\evec_L$ and $n_S=0$ scattered photons with wave
vector $\kvec_S$ and polarization $\evec_S$. The final state has one
laser photon less but one scattered photon (Stokes scattering). The
vector field can thus be written as a superposition of an incoming and
a scattered plane wave, $\Avec=\Avec_L+\Avec_S$ with
\begin{equation}
\Avec_S=A_S^+\evec_S^* e^{-i\kvec_Sr}~,\qquad
\Avec_L=A_L^-\evec_L e^{i\kvec_Lr}~,
\end{equation}
where $A_S^+$ contains the creation operator for the scattered photon
and $A_L^-$ the annihilation operator for the laser photons (note that
these are not hermitian).

Since Raman scattering is a second order process in $\Avec$, the term
$H_{AA}$ has to be treated in first order perturbation theory. It is
therefore nonresonant and includes only intraband scattering.  The
matrix elements are given by
\begin{eqnarray}
M^{(1)}_{n_fn_i}(\qvec,\kvec) 
	&=& {1\over2}r_0\braket{n_f\kvec+\qvec}{\Avec^2}{n_i\kvec} 
		\nonumber\\
	&=& r_0 \expv{A_S^+A_L}\evec_S^*\evec_L\delta_{n_in_f}~,
\label{fstpb}
\end{eqnarray}
whereas $\expv{A_S^+A_L}$ denotes a matrix element involving the
initial and final state of the photon field and
$\qvec=\kvec_L-\kvec_S$ is the momentum transfer from the photon field
to the sample. For other values of $\qvec$, the matrix element
$(\ref{fstpb})$ vanishes.

The second term, $H_A$, produces resonances via second order
perturbation theory. It has the form
\begin{equation}
M^{(2)}_{n_fn_i}(\qvec,\kvec) = r_0 \expv{A_S^+A_L}
	\sum_{n_m} \Gamma^{(2)}_{n_fn_i;n_m}(\qvec,\kvec)
\label{sndpb}
\end{equation}
with the expression
\begin{eqnarray}
\Gamma^{(2)}_{n_fn_i;n_m}(\qvec,\kvec) & = &  \label{second}\\
\noalign{\hskip0.2\baselineskip}%
&&\hspace{-2.5cm}
	{\braket{n_f\kvec+\qvec}{\evec_S^*\pvec}{n_m\kvec+\kvec_L}
	\braket{n_m\kvec+\kvec_L}{\evec_L\pvec}{n_i\kvec}
	\over\epsilon_{n_i\kvec}-\epsilon_{n_m\kvec+\kvec_L}+\omega_L+i0}
	+ \nonumber \\
&&\hspace{-2.5cm}
   +    {\braket{n_f\kvec+\qvec}{\evec_L\pvec}{n_m\kvec-\kvec_S}
	\braket{n_m\kvec-\kvec_S}{\evec_S^*\pvec}{n_i\kvec}
	\over\epsilon_{n_i\kvec}-\epsilon_{n_m\kvec-\kvec_S}-\omega_S+i0}~.
\nonumber
\end{eqnarray}
Here, $\omega_L$ and $\omega_S$ are the frequency of the incoming and
scattered light, respectively. Note that the states in the sum above
are states of the sample only. We have used Bloch states with band and
crystal momentum indices. The wavevectors of light, $\kvec_L$ and
$\kvec_S$, can usually be neglected in the matrix elements of
Eq.~$(\ref{second})$ because $v_F\ll c$. For the same reason,
$\epsilon_{n_m,\kvec+\kvec_L}\approx\epsilon_{n_m,\kvec}$. Therefore, we
introduce the symbol $\Gamma^{(2)}_{n_fn_i;n_m}(\kvec)$ to denote
expression~$(\ref{second})$ with the light wavevectors set equal to zero.

If we now add the contributions of both terms in Eqs.~(\ref{fstpb})
and~(\ref{sndpb}) and introduce second quantization, we are left with
the {\it effective Hamiltonian}
\begin{equation}
H_R = r_0\expv{A_S^+A_L}\,\tilde\rho_\qvec
\label{effham}
\end{equation}
as perturbation Hamiltonian leading to Raman scattering. The effective
density operator $\tilde\rho_\qvec$ can be expressed in the form
\begin{equation}
\tilde\rho_\qvec = \sum_{n_f,n_i,\kvec} \gamma_{n_fn_i}(\kvec)\,
	c^+_{n_f,\kvec+\qvec} c_{n_i,\kvec}~.
\label{flucop}
\end{equation}
using fermionic creation and annihilation operators for Bloch electrons
as well as the nondiagonal Raman vertex
\begin{equation}
\gamma_{n_fn_i}(\kvec) = \evec_S^*\evec_L\delta_{n_fn_i} +
	\sum_{n_m} \Gamma^{(2)}_{n_fn_i;n_m}(\kvec)~.
\label{raver}
\end{equation}
If we are interested mainly in the low frequency region, say Raman
shifts below $50\,\meV$, no {\it real} interband transitions of
significant weight are possible. This can easily be seen from the band
structure (Fig.~2 of Ref.~\onlinecite{AndJep}). Therefore, we
introduce the (intraband) {\it Raman vertex}
$\gamma_n(\kvec)=\gamma_{nn}(\kvec)$.

We proceed by discussing a very important simplification of
$(\ref{second})$ (with $n_i=n_f=n$ and $\qvec=0$), the {\it effective
mass approximation}. Four different cases will be discussed. First,
the virtual intraband transition with $n_m=n_i$. In this case, up to
first order in $v_F/c$, we have $\braket{n_m\kvec}{\pvec}{n_i\kvec} =
\braket{n_i\kvec}{\pvec}{n_m\kvec}$ (remember that $n_i=n_f$) and
$\epsilon_{n_i\kvec}-\epsilon_{n_m\kvec}=0$. Then, it can be seen that
the contributions of virtual {\it intra\/}band transitions relative to
the contribution of virtual {\it inter\/}band transitions to
intermediate states are of the order of the Raman shift over the laser
frequency, i.e.\ $\omega/\omega_L\ll1$, and can therefore be
neglected. The second case are the virtual {\it inter\/}band
transitions involving bands which are much farther away from the FS
than the light frequency. Then, because of
$\abs{\epsilon_{n_i}-\epsilon_{n_m}}\gg\omega_L$, the light
frequencies $\omega_L$ as well as $\omega_S$ can be neglected in
$(\ref{second})$. The third case also involves virtual interband
transitions, but for bands at about the laser frequency above the
Fermi surface. Here, the scattering is resonant, and the spectra are
expected to depend strongly on the laser wavelength. One can try to
avoid this situation by using different laser lines. So we assume that
in the third case $\omega_L$ and $\omega_S$ also can be
neglected. Finally, the forth case consists of virtual interband
transitions to neighboring bands with $\Delta\epsilon\ll\omega_L$. In
this case, neglecting $\omega_L$ and $\omega_S$ is more difficult to
justify. We do it nevertheless and reach the approximate conclusion
that we can neglect the light frequencies in Eq.~(\ref{second}) and
can restrict the sum in $(\ref{sndpb})$ to all $n_m\not=n_i$. Then,
Eq.~(\ref{raver}) becomes completely equivalent to the expression for
the inverse effective mass from $\kvec\cdot\pvec$ theory and we can
write
\begin{equation}
\gamma_n(\kvec)={m\over\hbar^2} 
	\sum_{i,j} \evec^*_{S,i} 
	{\partial^2\epsilon_{n\kvec}\over\partial k_i\partial k_j}
	\evec_{L,j}
\end{equation}
i.e.\ the Raman vertex is equal to the inverse effective mass
contracted with the polarization vectors of the laser light and the
scattered light, respectively.

Therefore, using the term $H_R$ with the intraband Raman vertex
$\gamma_{n\kvec}=\gamma_{nn}(\kvec)$ in Eq.~(\ref{flucop}) as
perturbation to the Hamiltonian for $\Avec=0$ and treating this in
first order perturbation theory is, under the mentioned
restrictions, equivalent to taking into consideration both
terms $H_A$ and $H_{AA}$.\cite{AbrGen}

According to the LMTO calculations, for Y-123 and Y-124 there are
bands above a band gap between approximately $2\,\eV$ and $2.5\,\eV$
above the Fermi energy. These bands can present a problem with respect
to the former discussion, because they are almost resonant for typical
laser wavelengths like $514.5\,{\rm nm}$. The same is true for the
conduction bands, which extend until $2\,\eV$ above the Fermi
surface. Note that due to the strong on-site interaction at the Cu-$d$
orbitals, correlation effects are expected to be important in the
electronic structure. It is possible that at energies of the order of
$1\,{\rm eV}$ or more above the Fermi surface the picture of the
Hubbard bands is a better description of the band structure and may
explain the weak dependence of the Raman spectra on the laser
frequency observed for laser frequencies in the visible range. The
band structure shows many bands at about the laser frequency below the
Fermi energy. These should yield resonant contributions to the Raman
efficiency.

Because the Raman vertex $\gamma_\kvec$ is, in the given
approximation, the second derivative of the energy with respect to
$\kvec$, the $A_{2g}$ component for tetragonal crystals vanishes in
this version of the theory ($A_{2g}$ is the symmetry of an
antisymmetric tensor). If one considers once more the effects of a
nearby resonance, it can be easily seen that the Raman tensor does not
have to be symmetric. This stresses again the questionability of the
effective mass approach if the scattering is resonant.


\subsection{The scattering efficiency}

Using the effective mass approach, we arrived at the effective
Hamiltonian $(\ref{effham})$ with the effective mass determining the
Raman vertex. This effective Hamiltonian is linear in
$\Avec_L\cdot\Avec_S$. The derivation of the scattering efficiency
using linear response theory is now a straightforward task.

The first step is finding a relation between the Raman efficiency and
a dynamical structure factor of the sample. Then, in a next step, the
fluctuation-dissipation theorem is used to connect the dynamical
structure factor to the imaginary part of a susceptibility, in our
case the {\it Raman susceptibility}.

To establish the relation to the dynamical structure factor, we add
the time evolution factor $e^{-i\omega t}$ to the effective
Hamiltonian $(\ref{effham})$ and use the golden rule to find the
transition rate from a state $i$ to a state $f$ of the sample. Then, we
sum over all final states $f$ of the sample and do a thermal averaging
over the initial states $i$. The transition rate from a state with
$n_L\equiv n_{\kvec_L\evec_L}$ laser photons and no scattered photon
to a state with $n_L-1$ laser photons and $n_S\equiv
n_{\kvec_S\evec_S}=1$ scattered photon at a temperature $T$ is given
by the expression
\begin{equation}
\Gamma^T(\kvec_L,\evec_L;\kvec_S,\evec_S) =
{2\pi\over\hbar} r_0^2\cdot \absii{\expv{A_S^+A_L}}\cdot
	\tilde S^T(\qvec,\omega)
\end{equation}
(the superscript $T$ denotes temperature dependency) whereas
\begin{eqnarray}
\tilde S^T(\qvec,\omega) &=& \nonumber\\
&&\hspace{-1cm}\sum_{i,f}{e^{-\beta E_i}\over\cal Z}
\absii{\braket{f}{\tilde\rho_\qvec}{i}} \delta(E_f-E_i+\hbar\omega)
\end{eqnarray}
is a {\it generalized dynamical structure factor} (of the sample!).
The partition function is denoted by ${\cal Z}$, and $\beta$ is the
inverse temperature.  Now, we sum over all final states in a certain
region $d\Omega\,d\omega_S$ of $k$-space around $\kvec_S$ and
normalize to the incoming flux $\hbar cn_L$. This yields the
expression
\begin{equation}
{d^2\sigma\over d\Omega\,d\omega}(\qvec,\omega) =
{\omega_S\over\omega_L} r_0^2 \tilde S^T(\qvec,\omega)
\label{rameff}
\end{equation}
for the differential cross section $d^2\sigma/d\Omega d\omega$ for a
given Raman shift $\omega$ and a given momentum transfer $\qvec$. This
differential cross section is proportional to the scattering
volume. When performing the calculation for a scattering volume equal
to unity, $\sigma$ becomes the commonly used {\it Raman scattering
efficiency}.

Finally, one can define a linear response function, the {\it Raman
susceptibility}
\begin{equation}
\chi_{\rm Raman}(\qvec,t) = {i\over\hbar}
\mathop{\rm Tr}\{{\cal Z}^{-1}e^{-\beta H_0}
[\tilde\rho_\qvec(t),\tilde\rho_{-\qvec}(0)]\}
\label{RamSus}
\end{equation}
and its Fourier-transformed $\chi_{\rm Raman}(\qvec,\omega)$. 
To relate the imaginary part of this quantity to the structure
function $\tilde S^T(\qvec,\omega)$, we use the
fluctuation-dissipation theorem. The result is
\begin{equation}
\tilde S^T(\qvec,\omega) = -{1\over\pi}(1+n_\omega)
	\Im\chi_{\rm Raman}(\qvec,\omega)
\label{fdt}
\end{equation}
with the Bose factor $n_\omega$. 

Equations $(\ref{rameff})$ and $(\ref{fdt})$ relate the Raman
efficiency directly to the imaginary part of the Raman
susceptibility. This evaluation of the Raman susceptibility shall be
given separately, {\rm (i)} in Sec.~\ref{sec:stheory} for the
superconducting phase and Raman shifts of the order of the gap, and
{\rm (ii)} in Sec.~\ref{sec:ntheory} for large Raman shifts in the
superconducting phase and for the normal phase.

\section{Theory: Superconducting phase}
\label{sec:stheory}

As pointed out in Ref.~\onlinecite{KleDie}, the Raman susceptibility
due to pair-breaking and including screening is given by a
polarization-like bubble made of a renormalized Raman vertex
$\Lambda_\kvec$, a Raman vertex $\gamma_\kvec$, and in between two
Green's function lines for Bogoliubov quasiparticles
(Fig.~\ref{screening}a).  The vertex renormalization includes
corrections for Cooper-pair-producing attractive interaction as well
as the repulsive Coulomb interaction, the Dyson equation for the
vertex $\Lambda_\kvec$ in the limit $\qvec\to0$ is given by Fig.~13 in
Ref.~\onlinecite{KleDie}.

To show more clearly the effect of screening, we write the equation
for the Raman susceptibility as given in Fig.~\ref{screening}b
and~\ref{screening}c. Figure~\ref{screening}b (with $a=\gamma_\kvec$
and $b=\gamma_\kvec$) shows the unscreened susceptibility
$\chi_{\gamma\gamma}$ given by a bare polarization bubble with two
Raman vertices $\gamma_\kvec$ and the contraction of a BCS-like ladder
sum with two Raman vertices. Therefore, $\chi_{\gamma\gamma}$ includes
the attractive Cooper-pair-producing interaction. We include Coulomb
screening by virtue of a RPA-like sum given in
Fig.~\ref{screening}c. The effect of screening on the electronic Raman
scattering can now easily be seen.\cite{AbrGen} If we denote by
$\chi_{ab}$ a bubble, renormalized by pairing interaction, with
vertices $a$ and $b$ at the ends as in Fig.~\ref{screening}b, the
RPA-chain can be easily summed up (see Fig.~\ref{screening}c) yielding
\begin{equation}
\chi_{\rm Raman}(\qvec\to0,\omega) = \chi_{\gamma\gamma}(\omega) 
	- {\chi_{\gamma 1}^2(\omega)\over\chi_{11}(\omega)}~,
\label{chir}
\end{equation}
where terms of order $\qvec^2$ have been dropped. In
Eq.~$(\ref{chir})$ we have used the fact that
$V_\qvec/(1-\chi_{11}V_\qvec)$ equals
$-\chi_{11}^{-1}(1-1/\varepsilon)$, and the factor $(1-1/\varepsilon)$
is $1+O(\qvec/q_{TF})^2$.

Without taking into account Coulomb interaction, the Green's functions
have a well-known massless pole (Goldstone mode) which is a
consequence of the breaking of gauge symmetry in the superconducting
phase.\cite{And} Coulomb interaction makes this pole acquire a finite
mass (which can be shown to correspond to the plasma frequency), so if
we correctly include Coulomb screening (not done in
Ref.~\onlinecite{BraCar}) we no longer have a Goldstone mode, but a
massive Anderson-Bogoliubov mode. This mode has the energy 
$\hbar\omega_p$ ($\omega_p$ is the plasma frequency) at the ${\it\Gamma}$
point and is therefore negligible for the low energy behavior of the
Raman spectra.

The susceptibilities $\chi_{ab}$ in Fig.~\ref{screening}b are like a
ladder sum contracted with vertices $a_\kvec$ and $b_\kvec$ and can be
written as a sum
\begin{equation}
\chi_{ab}(\qvec{=}0, \omega) =
	\sum_\kvec a_\kvec b_\kvec \lambda_\kvec(\omega)
\end{equation}
which involves the Tsuneto function\cite{Tsu}
$\lambda_\kvec(\omega)$. For small values of $\qvec$ (compared to the
inverse coherence length $\xi$ and the Fermi wave vector $k_F$), the
attractive interaction does not have to be taken into account in the
summation of the ladder, and the Tsuneto function is given simply by a
unmodified bubble and can be evaluated easily to be
\begin{eqnarray}
\lambda_\kvec(\omega) &=& {\Delta^2_\kvec\over E^2_\kvec}
	\tanh\left(E_\kvec\over 2T\right)\times\nonumber\\
&&\times\left({1\over2E_\kvec+\omega+i0}+{1\over2E_\kvec-\omega-i0}\right)~.
\label{tsufct}
\end{eqnarray} 
Equation~$(\ref{tsufct})$ involves the gap function $\Delta_\kvec$
(which depends on the temperature) and the quasiparticle dispersion
relation $E_\kvec^2=\xi_\kvec^2+\Delta_\kvec^2$ with
$\xi_\kvec^2=(\epsilon_\kvec-\epsilon_F)^2$. The constants $\hbar$ and
$k_B$ have been set equal to $1$.  As already mentioned, vertex
corrections due to the pairing interaction are neglected. This
approximation is valid for $q\ll\xi^{-1},k_F$
(Ref.~\onlinecite{DevEinPRB}) and $\omega\ll\omega_p$, because the
Anderson-Bogoliubov pole at the plasma frequency need no longer be
included.

A first and very important fact in the expressions above is that they
contain only the absolute square of the gap function, i.e.\ Raman
scattering is {\it not phase sensitive}, and consequently cannot
distinguish between a strongly anisotropic $s$ gap
$\left|\dxiiyii\right|$ and a $\dxiiyii$ gap.


In the preceding calculation of the unscreened correlation functions
$\chi_{ab}$, we have neglected impurity scattering as well as
scattering between quasiparticles (collisionless regime). In isotropic
$s$-wave superconductors at $T=0$ and for Raman shifts
$\omega\ll2\Delta$, it is perfectly reasonable to neglect impurity
scattering, because in this regime pair breaking is not
possible.\cite{AndImp} Also, the scattering between quasiparticles can
be neglected because their density is very small for small
temperatures $T\ll T_c$. For $d$-wave superconductors this is no
longer true. The effect of impurities will be discussed in the next
subsection, whereas a discussion about scattering between
quasiparticles can be found in Sec.~\ref{sec:ntheory}.

The second term of $(\ref{chir})$, representing screening, vanishes if
the average of $\gamma_\kvec\cdot\lambda_\kvec$ does. The Tsuneto
function is fully symmetric, i.e.\ has $A_{1g}$ ($D_{4h}$ group) or
$A_g$ ($D_{2h}$) symmetry regardless of gap symmetry. As a
consequence, the screening term vanishes unless the Raman vertex has
the same symmetry as the crystal. In the tetragonal case,
$A_{1g}$-like vertices are screened, but $B_{1g}$- and $B_{2g}$-like
are not.  This is different for orthorhombic HTSC of the YBCO-type. In
this case the Tsuneto function has $A_1$ symmetry, and the same is
true for the $\dxiiyii$-like component of the mass ($B_{1g}$ of
$D_{4h}$ group, $A_g$ of $D_{2h}$).  Consequently, in these
orthorhombic crystals, the $B_{1g}$ component is also screened. This
discussion is also applicable to BISCO, but with interchanged roles of
$B_{1g}$ and $B_{2g}$ modes because of the different orientation of
the crystallographic unit cell with respect to the Cu-O bonds.

In tetragonal systems, the $B_{1g}$ component of the Raman vertex has
nodes at the same position as the gap function. This has severe
consequences for the low-energy part of the spectra.\cite{DevEinPRB}
In two dimensions, the existence of the nodes of the gap function in
the case of a $\dxiiyii$ gap results in a linear density of states at
low energies. If the vertex has a finite value in this region, the
imaginary part of the Raman susceptibility is also linear in the
frequency. If the vertex has a node, however, its magnitude squared
becomes quadratic with respect to the gap on the Fermi surface. This
causes two additional powers of the frequency to appear, the $B_{1g}$
component of the scattering efficiency is cubic at low
frequencies.\cite{DevEin} Two effects can alter this behavior: an
orthorhombic distortion and impurities.

In our calculations, we focus on a $\dxiiyii$-like gap function which
is only a function of the direction in $k$-space, but not of the
magnitude of $\kvec$, since the values of the gap functions
sufficiently far from the Fermi surface do not affect the results.  We
are using the same gap function for all bands involved.

\subsection{Effect of impurities}
\label{sec:stheory:imp}

In contrast to scattering at non-magnetic impurities in conventional
(isotropic) superconductors, the influence of impurity scattering
plays an important role for superconductors with anisotropic gaps and
its effect on the Raman spectrum is most pronounced for
superconductors which exhibit regions in $k$-space where the gap
almost or completely vanishes. It was shown\cite{BorHir,DevImp} that
in the case of $d$-wave pairing, impurity scattering can be described
by extending the nodal points on the 2D FS to small finite regions
with vanishing gap. This causes a nonvanishing density of states at
the Fermi energy. For anisotropic $s$-wave pairing the gap anisotropy
becomes smeared out leading to an increase of the minimum gap value
$\Delta_{\rm min}$. In the case of a $\abs{\dxiiyii}$ gap, this
minimum gap increases monotonically with the impurity concentration
$n_i$ for small values of $n_i$.

The renormalization of the gap function by the presence of impurities
causes an additional contribution, which is linear in the Raman shift
$\omega$ for small Raman shifts $\omega$, in the Raman
spectra.\cite{DevImp} This has consequences for the $B_{1g}$ spectrum
of a {\it tetragonal} crystal, which, according to the theory, has a
cubic $\omega$-dependence, because a linear frequency dependence is
added. As will be discussed in the next subsection, the
orthorhombicity of the YBCO compounds also causes a linear addition to
the cubic behavior of the $B_{1g}$ channel spectrum.

In the case of a $\abs{\dxiiyii}$-like, $A_g$ symmetry gap function
the impurity-induced minimal gap $\Delta_{\rm min}$ causes an
excitation-free region to show up in the electronic Raman spectrum
below a Raman shift of $2\Delta_{\rm min}$.

\subsection{Effect of orthorhombic distortion}
\label{sec:stheory:odist}

As already mentioned, orthorhombic distortions, i.e.\ deviations
from the tetragonal symmetry, have a different effect on Y-123 and on
Bi-2212. Consider the $B_{1g}\,(D_{4h})$ component of the inverse mass
tensor in a {\it tetragonal} high-$T_c$ superconductor with a
$\dxiiyii$-like gap. The $B_{1g}\,(D_{4h})$ mass has its nodes in
directions diagonal to the axes of the copper planes; the same is true
for the gap function. As mentioned above, this results in the
$\omega^3$-dependence of the Raman efficiency for $B_{1g}\,(D_{4h})$
scattering, in contrast to the $\omega$-dependence predicted for
$A_{1g}$ and $B_{2g}$ scattering.  Let us now consider the
orthorhombic distortion present in Y-123. The zeros of the
$B_{1g}\,(D_{4h})$ mass shift because there are no longer mirror
planes through the $(110)$ axes. For this reason, the low-energy part
of the spectrum acquires a linear component in addition to the
$\omega^3$ component of the $D_{4h}$ case.

In Bi-2212 the situation is different because the orthorhombic
crystallographic cell is rotated by $45^\circ$ with respect to the
$a$- and $b$-axes: the orthorhombic distortion preserves the mirror
planes $[a\pm b,c]$.  Consequently, the $B_{1g}$ zeros stay at the
same position, the low-energy efficiency acquires no linear component.

\subsection{Effect of multilayers}
\label{sec:stheory:mlayer}

In systems with one layer of Cu-$\rm O_2$ planes per unit cell there
is only one sheet of Fermi surface and the mass fluctuations are
essentially intraband mass fluctuations, which are very sensitive to
the scattering polarizations. The scattering related to the average
mass is fully screened. The simplest $A_{1g}\,(D_{4h})$ scattering is
related to a Raman vertex of the form $\cos4\phi$ symmetry while
$B_{1g}\,(D_{4h})$ scattering is obtained for a $\cos2\phi$ vertex.
In multilayer systems, interband fluctuations between the various
sheets FS are also important. The lowest component of such
fluctuations corresponds to different {\it average} masses in each FS
sheet. Such fluctuations do not depend on the scattering polarizations
and lead to unscreened scattering of $A_g$ symmetry.

\subsection{Effect of sign change of $\gamma_\kvec$
on the Fermi surface}
\label{sec:stheory:signch}

The behavior of the Raman vertex near the Fermi surface, especially
its sign, is crucial for the scattering efficiency and, in particular,
for the effect of screening. {\it Antiscreening}, i.e.\ an {\it
enhancement} of the scattering efficiency by screening, can occur if
the Raman vertex changes sign on the Fermi surface. This can be seen
by considering the screening part
\begin{equation}
\Im\chiScr = - \Im{\chi_{\gamma1}^2\over\chi_{11}}
\label{scrpart}
\end{equation}
of the Raman susceptibility. A positive value of $\Im\chiScr$ enhances
the efficiency, i.e.\ corresponds to antiscreening.

To show how antiscreening arises, we first write the screening term
$\Im\chi_{\rm scr}$ in terms of the real and imaginary parts
$\Relam\equiv\Re\lambda$ and $\Imlam\equiv\Im\lambda$ of the
Tsuneto function and the Raman vertex $\gamma$ as
\begin{equation}
\Im\chiScr={
{\expv{\gamma\Relam}}^2\expv{\Imlam}-{\expv{\gamma\Imlam}}^2\expv{\Imlam}
-2\expv{\gamma\Relam}\expv{\gamma\Imlam}\expv{\Relam}\over
{\expv{\Relam}}^2+{\expv{\Imlam}}^2}~.\label{screefo}
\end{equation}
The imaginary part of the Tsuneto function $\Imlam$ is a positive
$\delta$-function. Consequently, the quantity $\expv{\Imlam}$ is a
positive function of the Raman shift $\omega$. If $\gamma_\kvec$
changes sign in a region around the Fermi surface, it is possible that
$\expv{\gamma\Imlam}$ changes sign as a function of $\omega$, i.e.\
has a zero. At the position of this zero, the second and the third
term in the numerator of $(\ref{screefo})$ vanish. The first term,
${\expv{\gamma\Relam}}^2\expv{\Imlam}$, is positive and can become
dominant in Eq.~(\ref{screefo}).  In this case antiscreening
results. In the Appendix~A will be shown that antiscreening is
particularly sensitive to the sign of the Raman vertex on parts of the
Fermi surface around the directions of the nodes of the gap function
$\Delta_\kvec$.

%
\section{Theory: Normal phase}
\label{sec:ntheory}

In the normal phase, the exact mechanism which produces a finite Raman
intensity almost constant over a broad frequency and temperature
range, is not known. Therefore, we assume some scattering mechanism,
which implies a finite lifetime of the quasiparticles. Candidates for
this scattering are the quasiparticle-quasiparticle scattering in
Marginal Fermi Liquid theory\cite{Var} (MFL), impurity
scattering\cite{HirWoe} or scattering due to spin
fluctuations.\cite{QuiHir} A self energy with non-vanishing imaginary
part yields a susceptibility of the form
\begin{equation}
\chi_{ab}(\qvec{=}0, \omega)=\sum_k a_\kvec b_\kvec\nu_\kvec(\omega)
\end{equation}
with the relaxation kernel (the function $f'$ is the derivative of the
Fermi function with respect to the energy)
\begin{equation}
\nu_\kvec(\omega) = -f'(\xi_\kvec) {i\Gamma_\kvec\over\omega+i\Gamma_\kvec}
\end{equation}
and its imaginary part
\begin{equation}
\Im\nu_\kvec(\omega) = -f'(\xi_\kvec) 
{\omega\Gamma_\kvec\over\omega^2+\Gamma_\kvec^2}~.
\label{imnu}
\end{equation}
This can easily be seen by evaluating a bubble with two Greens function
lines for quasiparticles with an imaginary part $\Gamma_\kvec$ of the
self energy.

Note that in the superconducting phase for Raman shifts larger than
$\sim\!\Delta$, the relaxation effects described by (\ref{imnu}) are
also of importance. The relevant relaxation kernel in this case is
\begin{equation}
\nu_\kvec(\omega) = -f'(E_\kvec) {\xi_\kvec^2\over E_\kvec^2} 
	{i\Gamma_\kvec\over\omega+i\Gamma_\kvec}~,
\label{drude}
\end{equation}
where $\xi_\kvec^2=(\epsilon_\kvec-\epsilon_F)^2$.  

To describe the constant background in the Raman spectra in the normal
phase, one has to adopt the quasiparticle scattering rate of the MFL
theory\cite{Var,VarII}
\begin{equation}
\Gamma_\kvec(\omega) \sim \max(\alpha T, \beta\omega)~.
\label{mfl}
\end{equation}
In order to evaluate the real part of $\nu_\kvec$ using causality
arguments, and to prevent divergences, we introduce a high-frequency
cutoff $\omega_C$.  Note that the nearly antiferromagnetic Fermi
liquid\cite{MMP,BarPin} (NAFL) and also the nested Fermi
liquid\cite{RuvVir} (NFL) yield a very similar quasiparticle
scattering rate. The former can also provide a mechanism, which
accounts for $\dxiiyii$ pairing.  Similar results are obtained with
Luttinger liquid based results.\cite{CarStr}

Equation~(\ref{mfl}) yields a scattering continuum which is constant
for frequencies smaller than $\min(\alpha T/\beta,T)$ and for
frequencies larger than the temperature $T$, but with different
intensities. In the first case, $\Gamma_\kvec$ is proportional to the
temperature, i.e.\ $\Im\chi\sim\omega/T$. Multiplying by the Bose
factor $1+n_\omega\sim T/\omega$ a constant is found. In the second
case, $\Gamma_\kvec\sim\omega$, and, consequently, $\Im\chi={\rm
const}$. The Bose factor is also constant and one is left with a
constant Raman intensity. Note that in the first case, $\Im\chi$
cancels the $\omega$- and $T$-dependence of the Bose factor. It has
been shown,\cite{DonKir,TZhou,VarII} that ${\rm Y}{\rm Ba}_2{\rm
Cu}_4{\rm O}_8$ does not exhibit this behavior. This has been
attributed to the breakdown of MFL theory for not optimally doped
cuprates.\cite{VarII} Actually, in this case the spectra are nearly
temperature independent {\it after} dividing them by the Bose
factor. We shall address this question once more at the end of this
section.

To discuss quasiparticle-quasiparticle (qp-qp) scattering, and its
influence on electronic Raman scattering, we start with the case of a
$\dxiiyii$ gap. Suppose the nodes of this gap have a width $\delta_0$
in $k$-space on the Fermi surface due to impurity scattering. We use
the model of Eq.~(\ref{drude}) with a quasiparticle scattering rate
$\Gamma_\kvec$ independent of $\kvec$ and discuss first the case
$T=0$. Then it can be seen that the contribution of qp-qp scattering
to the imaginary part of the Raman susceptibility $(\ref{RamSus})$ for
low frequencies $\omega\ll\Delta_{\rm max}$ is proportional to the
Drude-like factor $\omega\Gamma/(\omega^2+\Gamma^2)$ (which is, for
small $\omega$ and low temperatures $T<\omega$, linear in $\omega$ if
$\Gamma=\hbox{const}$ (semiconductors) or
$\Gamma\sim\max(\omega^2,T^2)$ (FL), but constant as a function of
$\omega$ if $\Gamma\sim\max(\omega,T)$ (MFL). In the tetragonal case,
it is also proportional to the density of states at the Fermi surface
and in the case of $A_{1g}$ and $B_{2g}$ polarizations to the width
$\delta_0$, and in the case of $B_{1g}$ to the third power
$\delta_0^3$ of the width $\delta_0$. The discussion for BISCO is
analogous with the exception that $B_{1g}$ and $B_{2g}$ exchange their
role.

Finite, but small temperatures $T\ll\Delta_{\rm max}$ have the effect
of enlarging the widths $\delta_0$ linearly in temperature, i.e.\ the
temperature dependence of the contribution from qp-qp scattering is
proportional to ${\rm const}+T$. Note that for $T\appgeq0$, the Bose
factor changes the linear-in-$\omega$ dependence to a constant.

For the anisotropic $s$ gap of the form $\abs{\dxiiyii}$ which
acquires a finite minimum gap $\Delta_{\rm min}$ due to the presence
of impurities,\cite{DevImp} the situation is different. The frequency
dependence is also given by the factor
$\omega\Gamma/(\omega^2+\Gamma^2)$ in addition to the Bose factor. But
the temperature dependence is different. For temperatures
$T\ll\Delta_{\rm min}$ smaller than the minimal gap, the density of
quasiparticles is proportional to $\exp(-\Delta_{\rm min}/kT)$, i.e.\
the contribution of qp-qp scattering to the Raman efficiency is
exponentially small. At $kT\approx\Delta_{\rm min}$, this exponential
dependence on $T$ crosses over to a power law.

The background electronic Raman spectrum in the normal phase is almost
independent of temperature for nearly optimally doped high-$T_c$
compounds only. In the overdoped and underdoped case, the materials
seem to show Fermi liquid-like behavior concerning the quasiparticle
scattering rate $\Gamma_\kvec$ (for small $\omega$).\cite{VarII,TZhou}
The temperature dependence of the scattering rate $\Gamma_\kvec$, as
defined in $(\ref{drude})$, has been measured\cite{HacNem} for
optimally doped and overdoped Bi-2212, and, especially in the case of
the $B_{2g}\,(D_{4h})$ mode, the optimally doped sample shows
$\Gamma=\alpha T$, whereas for the overdoped sample
$\Gamma=\alpha'T^2+\Gamma_0$. Therefore, the overdoped sample shows
properties of a normal Fermi liquid which are predicted by theory to
have $\Gamma\sim\max(\omega^2,T^2)$. The $B_{1g}\,(D_{4h})$ mode
result for the optimally doped sample yields the puzzling
quasiparticle scattering rate $\Gamma={\rm const}$.

\section{Experimental spectra}
\label{sec:expts}


The experimental determination of {\it absolute} Raman scattering
intensities is plagued by a number of difficulties (a reason why
usually ``relative units'' are found in the literature). The first is
related to the presence of elastically scattered light in the spectra,
in particular when non-ideal sample surfaces are involved. Depending
on the quality of the spectrometer this leads to contributions
extending typically, for the parameters of the present work, up to
$50\,\waven$ from the center of the laser line. These contributions
can be filtered out using a premonochromator or notch filters but, in
any case, Raman scattering measurements below $50\,\waven$ remain
difficult. The measurements discussed here have been performed by
comparison with the known efficiency of silicon after correcting for
differences in the scattering volumes. The procedure leads to errors
of about 50\%.


We use for comparison with the calculation the experimental data of
Krantz {\it et al.}\cite{KraCar} in the case of Y-123, and Donovan
{\it et al.}\cite{DonKir} in the case of Y-124. Our
Figs.~\ref{Y123exp} and~\ref{Y124exp} are taken from these
publications. In the case of Fig.~\ref{Y123exp} we have corrected a
scale error in the abscissa found in Ref.~\onlinecite{KraCar}. In the
case of Fig.~\ref{Y124exp} we have calculated the $A_{1g}$ component
from the experimental results for the $(x'x')$ and $(xy)$
polarizations.


The classification of the measured spectra according to irreducible
representations of the symmetry group of the crystal is performed with
the use of the Raman tensor $\hat R$ which is related to the Raman
efficiency through the expression $I\sim{|\evec_L\hat R\evec_S|}^2$,
bilinear in the Raman tensor.  In the calculations, the Raman tensor
does not appear explicitly, the inverse effective mass $\partial^2
E/(\partial k_i\partial k_j)$ playing its role. It is important to
note that the Raman efficiency as given by the theory
(Eqs.~(\ref{rameff}), (\ref{fdt}), and~(\ref{chir})) is bilinear in
the inverse effective mass of the Raman vertex (including the
screening part!), i.e. contains the same interferences as the approach
involving the Raman tensor. Note that the Tsuneto function $\lambda$
is fully symmetric. In the normal phase, the scattering kernel $\nu$
has been assumed to be the same for all scattering channels.

In most of the measurements of the Raman efficiency in orthorhombic
high-$T_c$ superconductors, an $A_{1g}$ component has been
given. Strictly, this irreducible representation does not exist in
$D_{2h}$ but only in $D_{4h}$. In orthorhombic crystals, the Raman
tensor contains two $A_g$ components which correspond to the $A_{1g}$
and $B_{1g}$ components of the tetragonal $D_{4h}$ case, and which are
not distinguishable in $D_{2h}$ because they transform in the same
way. Nevertheless, quantities can be constructed in the orthorhombic
case which correspond to the tetragonal $A_{1g}$ component.

One of these is $I^{(1)}=(I_{xx}+I_{yy})/2-I_{x'y'}$. Both, $I_{xx}$
and $I_{yy}$ contain $A_{1g}$ and $B_{1g}\,(D_{4h})$, and also an
interference term which cancels when $I_{xx}$ and $I_{yy}$ are added.
The $I_{x'y'}$ efficiency contains $B_{1g}$ and $A_{2g}\,(D_{4h})$. If
we assume that the antisymmetric component ($A_{2g}$ in $D_{4h}$) of
the Raman tensor $\hat R$ vanishes (i.e.\ $I_{xy}=I_{yx}$), $I_{x'y'}$
corresponds to tetragonal $B_{1g}$ and cancels the $B_{1g}$
contribution in $I_{xx}$ and $I_{yy}$. Provided that the $A_{2g}$
component of the Raman tensor vanishes, $I^{(1)}$ corresponds to the
$I_{A_{1g}}$ of the tetragonal case. Note that the antisymmetric
compoment $(R_{xy}-R_{yx})/2$ of the Raman tensor vanishes in the
effective mass vertex theory given in Sec.~\ref{sec:gtheory} because
of $\gamma_{xy}=\gamma_{yx}$ regardless of the symmetry of the
crystal, and also in the experiment in the case of tetragonal crystals
but not necessarily for orthorhombic crystals. The equality of
$I_{xy}$ and $I_{yx}$ in the calculation is an artifact of the theory.

A second possible construction for $A_{1g}$ is $I^{(2)}=I_{x'x'}-I_{xy}$.
The $I_{x'x'}$ efficiency contains $A_{1g}$ and $B_{2g}$
contributions. The interference term of these two contributions
vanishes in the tetragonal as well as the orthorhombic case.
Both, $B_{2g}\,(D_{4h})$ and $A_{2g}$ are contained in $I_{xy}$. But
if the $A_{2g}$ component of the Raman tensor vanishes, $I^{(2)}$ also
corresponds to the $I_{A_{1g}}$ of the tetragonal case.  In one of the
experimental works\cite{KraCar} a different method to extract the
$A_{1g}$ component was used.  Both of the expressions for $I^{(1)}$
and $I^{(2)}$ contain contributions of the $A_{2g}\,(D_{4h})$ Raman
tensor component. This component may be present in the experiment, but
not in the theory, a fact, that has to be kept in mind when comparing
the numerical results to the measurements. Note that the Raman
efficiencies in $(xy)$ and $(x'y')$ polarization configurations also
contain contributions from the antisymmetric part of the Raman tensor.
In view of these uncertainties in $A_{1g}$ we mainly focus in the next
section on the directly observable components of the Raman tensor.

We shall conclude this section by taking up again the question of the
validity of the effective mass approximation.  In the experiment, this
can be checked in two ways. First, via the dependence of the spectra
on the laser frequency which should make it possible to distinguish
the contributions to the Raman efficiency resulting from resonant and
non-resonant transitions, respectively.  The second way involves the
measurement of the $A_{2g}$ component of the mass. If the effective
mass approximation is valid, the Raman vertex should be symmetric
($\gamma_{xy}=\gamma_{yx}$), i.e.\ the $A_{2g}\,(D_{4h})$ component
should vanishes. A non-vanishing $A_{2g}$ component of the measured
scattering would cast doubts on the appropriateness of the effective
mass approximation.

\section{Numerical results and discussion}
\label{sec:numerics}


To carry out the numerical BZ and FS integrations, we employed a
tetrahedron approach.\cite{LehTau,JepAnd} The convergence of the
integrations was checked by using different meshes. In
Figs.~\ref{Y123} and~\ref{Y124}, the results of full BZ integrations
for Y-123 and Y-124, respectively, are plotted.  The corresponding
spectra obtained through FS integrations can be seen in
Ref.~\onlinecite{CarStr}. The Bose factor has not been included, hence
the results apply to zero temperature.  In both figures, the Raman
shift is given in units of the gap amplitude $\Delta_0$. Since the
calculated scattering efficiencies for BZ integrations, contrary to FS
integrations, are not only a function of the reduced frequency but
depend also weakly on the value of $\Delta_0$, we took for the
calculations $\Delta_0=220\,\waven$. This value of $\Delta_0$ falls in
the range of $\Delta_0$'s determined by Raman scattering and other
methods. The delta-function peaks in the Tsuneto function have been
broadened phenomenologically by introducing a finite imaginary part
$\Gamma=0.3\Delta_0$ of the frequency variable $\omega$.

Figures~\ref{Y123} and~\ref{Y124} display spectra for each of the
polarization configurations $(yy)$, $(x'x')$, $(xx)$, $(x'y')$, and
$(xy)$, as well as the symmetry component $A_{1g}\,(D_{4h})$ (defined
by $I_{A_{1g}}=I_{x'x'}-I_{xy}$), the unscreened intensities, the
screening part $(\ref{scrpart})$, and the total intensities, equal to
the difference between unscreened and screening parts. Note that the
$(x'y')$ configuration corresponds to the $B_{1g}\,(D_{4h})$ component
because of the vanishing of the $A_{2g}$ component in the theory.


We discuss first the results for Y-123.  The $A_{1g}$ component (in
the rest of this section we use tetragonal notation unless explicitly
stated) is subject to rather strong screening, however its unscreened
part is comparable to that of the $B_{1g}$ component.  The relation
between the unscreened and the screened (total) spectral weight of the
$A_{1g}$ component is about three. Nevertheless, the shapes of the
unscreened and the screened parts are the same and, consequently, {\it
there is almost no shift in the peak position due to screening}
(contrary to the results of Ref.~\onlinecite{DevEin}). The peak is
located almost exactly at $2\Delta_0$. Note that there is no
antiscreening in the $A_{1g}$ component. The low-energy part of all
$A_{1g}$ spectra (screened and unscreened) is linear, as predicted by
the theory.

As already mentioned, the $(x'y')$ component (equal to the $B_{1g}$
component in the non-resonant case) is almost four times stronger than
its screened $A_{1g}$ counterpart. The screening is very small, its
nonvanishing being an effect of the distorted tetragonality of the
crystal. There is, in this case, a very small amount of antiscreening
in the region below $2\Delta_0$. As in the case of the $A_{1g}$
component, the $(x'y')$ component peaks at almost exactly the
$2\Delta_0$ frequency shift. The low-frequency part has an
$\alpha\omega+\beta\omega^3$ frequency dependence, the linear part
arising from the distorted tetragonality, i.e.\ the fact that the
$B_{1g}$ mass does not vanish at exactly the same position on the
Fermi surface as the gap function does.

The efficiency of the peak in the $(xy)$ configuration (equal to the
$B_{2g}$ component in the non-resonant case) is also four or five
times smaller than that of the $A_{1g}$ peak. The $(xy)$ peak is
located at about $1.3\Delta_0$, as expected from the fact that in the
neighborhood of the region where the gap is large, the $B_{2g}$ mass
vanishes.  Consequently, the peak is not as sharp as in the former
cases and screening vanishes since these spectra correspond to a
nonsymmetric ($B_{1g}$) representation of the orthorhombic group
($D_{2h}$).

In the $A_{1g}$ and $(x'y')$ spectra there should be a small peak at
about $\omega=2\sqrt{\epsilon_{\rm vH}^2+\Delta_{\rm
max}^2}\approx3.9\Delta_0$ due to the van Hove singularity on the
$k_x$-axis near the $X$ point. The corresponding structure, however,
is very weak, and practically invisible in Fig.~\ref{Y123}. This is
not unexpected for a 3D calculation. These peaks appear strongly when
2D calculations are performed through BZ integrations.\cite{BraCar}


In general, the efficiencies in Y-124 (Fig.~\ref{Y124}) are about a
factor of three less than those for its Y-123 counterpart. Moreover,
the screening of the $A_{1g}$ component of Y-123 is much stronger than
that of Y-124. This may be, at least in part, due to the additional
chain band: The $(yy)$ component of Y-124 is less screened than the
$(yy)$ component of Y-123. At low frequencies, we correspondingly have
antiscreening even in $A_{1g}$, a fact which reveals a change of sign
of the effective mass on the Fermi surface (see
Sec.~\ref{sec:stheory:signch}). Due to this antiscreening, the peak
in the $A_{1g}$ spectrum is shifted from $2\Delta_0$ towards
approximately $1.6\Delta_0$. In contrast to the situation in Y-123,
the Y-124 spectra show clearly the influence of the van Hove
singularity on the spectra, as a small hump (vH) located near
$2\sqrt{\epsilon_{\rm vH}^2+\Delta_{\rm max}^2}\approx7\Delta_0$.  In
the $A_{1g}$ spectrum this hump is almost screened out whereas in the
$(x'y')$ spectrum it appears slightly increased by the influence of
antiscreening.


To compare these predictions with the experiment let us first focus on
the peak positions. The experimental results for Y-123
(Fig.~\ref{Y123exp}, lower part) clearly show that the position of the
$(yy)$, $(x'x')$ and $(xx)$ peaks is at about $300\,\waven$, whereas
the $(x'y')$ peak is located at $600\,\waven$, i.e.\ at twice the
frequency of the former. This fact is in sharp contrast with the
calculated spectra and has been at the center of the controversy
concerning the topic at hand.\cite{KCComm,DevRepl} It has been
suggested by Devereaux {\it et al.}\cite{DevEin,DevRepl} that the
$B_{1g}$ component peaks at $2\Delta_0$, and the $A_{1g}$ component
becomes shifted down to almost $\Delta_0$ by the screening. This
interpretation contradicts our numerical results which clearly suggest
that the influence of screening on the position of the $A_{1g}$ mode
is usually smaller.  The frequency renormalizations of phonons around
$T_c$ also seem to contradict the interpretation in
Refs.~\onlinecite{DevEin} and~\onlinecite{DevRepl}.  It has been
shown\cite{FriTho} that lowering the temperature of the sample in the
superconducting phase causes the $A_{1g}$ $435\,\waven$ phonon
(plane-oxygen, in-phase) to shift up in frequency and the $B_{1g}$
($D_{4h}$ notation) $340\,\waven$ phonon (plane-oxygen, out-of-phase)
to shift down. This, in turn, implies an amplitude of the gap
$2\Delta_0$ between $300\,\waven$ and $360\,\waven$ and is consistent
with our interpretation of the electronic Raman spectra with the
$A_{1g}$ peak at $2\Delta_0$.

Note that the $(yy)$, $(x'x')$ and $(xx)$ spectra do {\it not} contain
contributions of the $A_{2g}\,(D_{4h})$ antisymmetric component of the
Raman tensor while the $(x'y')$ component does. So, the experimental
results may suggest that the shift of the position of the $(x'y')$
spectrum with respect to the peak position of the other spectra is
due to resonance effects. The $(xy)$ spectrum is also influenced by
the $A_{2g}$ component. It is difficult to determine its peak position
from Fig.~\ref{Y123}, but it seems to be located at the same position
as that of the $(yy)$, $(x'x')$ and $(xx)$ configurations. The
calculation predicts it to be located at about $1.3\Delta_0$, the
shift to $2\Delta_0$ can also be attributed to the existence of an
$A_{2g}$ component, like in the case of the $(x'y')$ configuration.


To compare the relative intensities of the spectra with different
polarizations, we refer to Table~\ref{peakh}, which lists them
together with the corresponding absolute intensities, both at the peak
position. The detailed results of our FS integration have already been 
reported earlier.\cite{CarStr} We begin with Y-123 (upper panel in
Table~\ref{peakh}) and compare BZ integration results to the
experimental ones.  With the possible exception of the $A_{1g}$
component (and the $(x'x')$ component, which is very similar to
$A_{1g}$), the agreement is rather good. The deviation of the $A_{1g}$
component may be attributed to screening, which is very sensitive
to sign changes and other details of the Raman vertex near the Fermi
surface (such as details of the band structure and especially the
exact position of the Fermi energy). 

The second compound, Y-124 (lower panel in Table~\ref{peakh}), also
shows reasonable agreement between the results of the BZ integration
and the experiment. However, we also have problems with the $A_{1g}$
component, as we did for Y-123.


The measured absolute intensities agree particularly well with the
calculations in the case of Y-123. With the exception of $A_{1g}$, the
discrepancy between theory and experiment is only a factor of two,
which can easily be related to the difficulties in measuring absolute
scattering cross sections.  In the case of Y-124, the discrepancy is a
bit larger, but a factor of four can still be considered good.
We should also keep in mind that resonances of $\omega_L$ or
$\omega_S$ with virtual interband transitions are expected to enhance
the simple effective mass Raman vertex, a fact which could also
explain why the measured scattering efficiencies are usually larger
than the calculated ones.


We close the discussion of the numerical results with a remark about
the Fermi surface integration.  For Y-124, the results of the former
correspond rather closely to the results from the BZ integration. The
situation is different for Y-123.  Here, the $(xx)$ peak height is
almost a factor four larger in the FS integration than in the BZ
integration. This is likely to result from the close proximity of the
van Hove singularity to the FS in the case of Y-123 ($25\,\meV$), as
compared to Y-124 ($110\,\meV$).


To verify the predictions related to the effect of orthorhombic
distortions as discussed in Sec.~\ref{sec:stheory:odist}, we performed
a fit of the function $\alpha\omega+\beta\omega^3$ to the
low-frequency part of the $B_{1g}$ data for Y-123 reported in
Ref.~\onlinecite{KraCar} and Ref.~\onlinecite{Hacetal} as well as for
Bi-2212 (taken from Ref.~\onlinecite{Staetal}) and to the results of
our numerical calculations for \hbox{Y-123}. The ratios of the cubic
vs.\ the linear part (at $\omega=300\,{\rm cm}^{-1}$) of the fit to
the low-frequency efficiency are given in Table~\ref{peakh}.

Both measurements for Y-123 agree in their large linear part, which
should be due mainly to the lack of exact tetragonality and the
presence of impurities. The results of the BZ integration show a
smaller linear part, because they do not take into account the
influence of impurities. Finally, the result for Bi-2212 is completely
different from the former results for Y-123. The linear part almost
vanishes, in agreement with the preceding discussion.

\section{Conclusions}
\label{sec:concl}

In spite of the striking ability to predict not only general features
of the observed spectra but also their peak intensities, our
calculations are not able to predict the relative positions of the
$A_{1g}$ and $B_{1g}$ peaks. According to Figs.~\ref{Y123}
and~\ref{Y124} the $A_{1g}$ spectrum should peak only slightly below
$2\Delta_0$ while $B_{1g}$ should peak at $2\Delta_0$. The
experimental data of Figs.~\ref{Y123exp} and~\ref{Y124exp}, however,
indicate that the $B_{1g}$ spectra peak nearly at twice the frequency
of $A_{1g}$. Since the observed $A_{1g}$ peak is considerably sharper
than that of $B_{1g}$, we may want to assign the $A_{1g}$ peak to
$2\Delta_0$. Our calculations show that it is impossible to reproduce
both peak frequencies with a simple gap of the form
$\Delta_0\cos2\phi$ where $\phi$ is the direction of the
$\kvec$-vector. A reasonable fit was obtained in
Ref.~\onlinecite{KraCar} with a two-dimensional FS which did not take
into account the chain component and assigned $d$- and $s$-like gaps
to the two bonding and antibonding sheets of the FS of the two planes
in an {\it ad hoc} way. Within the present 3-dimensional band
structure the FS cannot be broken up into bonding and antibonding
plane and chain components since such sheets are interconnected at
general points of $k$-space. It is nevertheless clear that there is no
reason why the gap function should be the same in the various sheets
for a given $\kvec$-direction. Thus the remaining discrepancy in the
peak positions between theory and experiment could be due to a more
complicated $\Delta_{n\kvec}$ than a simple $\Delta_0\cos2\phi$ used
here. Another possible source of this discrepancy is scattering
through additional excitations of a type not considered here (e.g.\
magnetic excitations) contributing to and broadening the $B_{1g}$
peak.

A BCS-like theory, which involves an attractive pairing potential as
well as the repulsive Coulomb potential and uses an anisotropic
$\dxiiyii$-like gap function in connection with the effective mass
approximation used in the calculation of the absolute Raman scattering
efficiencies yields result which are in significant agreement with the
experimental spectra. One exception, the peak positions of the
$A_{1g}$ and the $B_{1g}$ components, remains unexplained.  The theory
predicts them to be both located near $\omega=2\Delta_0$, but the
experiment shows the peak in $B_{1g}$ at almost twice the frequency of
the peak in $A_{1g}$.  The weak $B_{2g}$ spectrum agrees in intensity
and peak position with calculations for a $\dxiiyii$-like gap.  The
results of other experiments, involving the temperature dependence of
phonon frequencies,\cite{FriTho} suggest that the $A_{1g}$ peak
position corresponds to the gap amplitude $2\Delta_0$. The shifting of
the $B_{1g}$ peak towards higher frequencies may have an origin
different from the mass-fluctuation-modified charge-density
excitations described in the theoretical part of this paper but could
also be due to a multi-sheeted gap function, more complicated than the
simple $\dxiiyii$-like $\Delta_0\cos2\phi$ gap assumed in our
calculations. The initial variation of the $A_{1g}$ and $B_{1g}$
scattering efficiencies vs.\ $\omega$ are linear as expected for that
gap. The $B_{1g}$ symmetry becomes $A_g$ in the presence of the
orthorhombic distortion related to the chains.  Consequently, the
scattering efficiency at low frequency is not proportional to
$\omega^3$ but should have a small linear component which is found
both in the calculated and the measured spectra. In the corresponding
spectrum of Bi-2212, with and orthorhombic distortion along $(x+y)$,
the $B_{1g}$ ($D_{4h}$) excitations also have a nonsymmetric $B_{1g}$
($D_{2h}$) orthorhombic character. Consequently, for small $\omega$ no
component linear in $\omega$ is found in the measured spectra. 

We have performed our calculations using either BZ or FS integration.
In the case of Y-124 the spectra so obtained are very similar. For
Y-123 quantitative differences appear; they are probably related to
the presence of a van Hove singularity close to the FS. These
singularities appear as weak structures in the calculated spectra, as
expected for a 3D band structure.

\acknowledgments 

We thank Jens Kircher for providing us with the LMTO band structures.
One of us (TS) also would like to thank his colleagues at the MPI for
numerous discussions on Raman scattering and superconductors.  Thanks
are specially due to Igor Mazin for a critical reading of the
manuscript.

\appendix
\section{Antiscreening and the sign of the Raman vertex}
\label{sec:appa}

In Sec.~\ref{sec:stheory:signch}, we pointed out that the effect of
antiscreening results from sign changes of the Raman vertex
$\gamma_\kvec$, i.e.\ the inverse effective mass, on the Fermi
surface. In this appendix we present a different proof using a power
expansion of $\gamma_\kvec$.

For very low frequencies $\omega\ll\Delta_0$, only the regions around
the node directions of the gap function (we assume a
$d_{x^2-y^2}$-like gap) contribute to the susceptibility. We focus on
a specific node of the gap function and define the point $\kvec_0$ as
the point of intersection of the node line of the gap function and a
specific sheet of the Fermi surface. Then we introduce an orthogonal
coordinate system $\{k,k_\perp\}$ in the $a$-$b$-plane in $k$-space
with the origin at $\kvec_0$, rotated in such a way that the
$k_\perp$-axis is perpendicular to the node line (i.e.\ tangent to the
Fermi surface).  We write the Raman vertex as a series
$\gamma(\kvec)=\sum_i\gamma_i(k_\perp/k_c)^i$ (where $k_c$ is a
cutoff), using the assumption that $\partial\gamma/\partial k=0$. This
approximation is justified since contributions to the Raman
susceptibility mostly arise from a narrow region around the Fermi
surface). We write Eq.~(\ref{screefo}) as
\begin{equation}
\Im\chiScr=\sum_i\Im\chi_{\rm Scr}^{(i)}~,
\end{equation}
whereas $\Im\chi_{\rm Scr}^{(i)}$ contains $i$-th powers in $k_\perp$
from the expansion of $\gamma_\kvec$. The first three terms in this
sum are
\begin{eqnarray}
\Im\chi_{\rm Scr}^{(0)} &=& -\gamma_0^2\expv{\Imlam}\nonumber\\
\Im\chi_{\rm Scr}^{(1)} &=& -2\gamma_0\gamma_1\expv{k_\perp\Imlam}\\
\Im\chi_{\rm Scr}^{(2)} &=& -2\gamma_0\gamma_2\expv{k_\perp^2\Imlam}\nonumber\\
&&\hspace{-1.5cm}-\gamma_1^2{\expv{\Imlam}({\expv{k_\perp\Imlam}}^2-
{\expv{k_\perp\Relam}}^2)
+2\expv{\Relam}\expv{k_\perp\Relam}\expv{k_\perp\Imlam}\over
{\expv{\Relam}}^2+{\expv{\Imlam}}^2}~.\nonumber
\end{eqnarray}
Now we investigate the behavior of $\expv{k_\perp^i\Imlam}$ in the low
frequency limit. We set $v_F=1$ and $\Delta_0=1$ (this represents a
simple change in scales). Then, $E_\kvec=k$ ($E_\kvec$ is constant as
function of $k_\perp$ by definition), $\Delta_\kvec=k_\perp$, and
therefore $\Delta_\kvec^2/E_\kvec^2=k_\perp^2/k^2$ and
$\Imlam\sim(k_\perp/k)^2\cdot\delta(2k-\omega)$ from
Eq.~(\ref{tsufct}). We perform a 2D BZ integration which has to be cut
off at a value proportional to $\omega$ in the $k_\perp$ integration
and find
\begin{eqnarray}
\expv{k_\perp^i\Imlam}&\sim&\int dk_\perp\,dk\,k_\perp^i\cdot
{k_\perp^2\over k^2}\delta(2k-\omega)\nonumber\\
&\sim&\omega^{-2}\int^\omega dk_\perp\,k_\perp^{i+2}
\sim\omega^{i+1}
\label{expvalue}
\end{eqnarray}
as the low frequency behavior. The same is true for the real part
$\expv{k_\perp^i\lambda'}$. Note that the proportionality constant in 
$(\ref{expvalue})$ is positive. This implies that $\Im\chi_{\rm
Scr}^{(i)}\sim\omega^{i+1}$ and 
\begin{eqnarray}
\Im\chiScr &=& -\left(\alpha_0\gamma_0^2\omega
	+\alpha_1\gamma_0\gamma_2\omega^2
	+(\alpha_2\gamma_0\gamma_2+\alpha_3\gamma_1^2)\omega^3
	\right.\nonumber\\
	&&\qquad\left.+O(\omega^4)\right)
\end{eqnarray}
with positive constants $\alpha_0$, $\alpha_1$ and $\alpha_2$. The
sign of $\alpha_3$ depends on the specific case.

For a tetragonal Fermi surface, we only have to focus on the $A_{1g}$
mode because the screening contributions to the other components
vanish by symmetry. The $A_{1g}$ symmetry implies $\gamma_1=0$ because
of the $\sigma_d$ symmetry operation (reflection at $\Gamma S$) which
transforms $\gamma_1k_\perp\to-\gamma_1k_\perp$. Then, the screening
term for low frequencies can be written as
\begin{equation}
\Im\chi_{\rm Scr}=-\left(\alpha_0\gamma_0^2\omega
	+\alpha_2\gamma_0\gamma_2\omega^3+\cdots\right)~.
\end{equation}
If there are nodes in the Raman vertex near the node of the gap
function, then $\gamma_0\gamma_2<0$, i.e.\ the screening term is
negative for very small $\omega$, but eventually crosses zero because
of the $\omega^3$ contribution. For large $\omega\appgeq2\Delta_0$,
the approximation $\expv{\gamma\lambda}\approx\expv\gamma\expv\lambda$
yields $\chiScr=-{\expv\gamma}^2\chi_{11}$, which is the screening
term for an isotropic Fermi surface with a scalar Raman vertex
$\expv\gamma$, and therefore negative. If the Raman vertex does not
show nodes near the node of the gap function, $\gamma_0\gamma_2$ is
larger than zero, and the screening term is not likely to change sign.

In the orthorhombic case of a weakly distorted tetragonality, the
$B_{1g}\,(D_{4h})$ zero in the Raman vertex may shift with respect to
the gap node. This can be described by a small $\gamma_0\not=0$ and
$\gamma_1\not=0$; $\gamma_2$ can be neglected. The low frequency
screening term then has the form
\begin{equation}
\Im\chi_{\rm Scr}=-\left(\alpha_0\gamma_0^2\omega
	+\alpha_1\gamma_0\gamma_1\omega^2
	+\alpha_3\gamma_1^2\omega^3\cdots\right)~.
\end{equation}
The first an the third term have almost $A_{1g}$ symmetry whereas the
second term has almost $B_{1g}\,(D_{4h})$ symmetry.  It can be shown
that the term proportional to $\omega^2$ vanishes if the other nodes
of the gap function are taken into account. Whether antiscreening
exists or not depends on the sign of $\alpha_3$.

It can be seen that for small values of $\gamma_0$, the antiscreening
can start already at very small Raman shifts as is the case in the
calculations, Fig.~\ref{Y124}, $B_{1g}$ panel.


\newpage 

\end{multicols}


\begin{table} 

\caption[(caption tab. 1)]{%
The ratio between the linear and cubic parts of the low energy Raman
efficiency in $B_{1g}\,(D_{4h})$ configuration of several high-$T_c$
compounds at a Raman shift of $\omega=300\,{\rm cm}^{-1}$.}
\label{tab1}

\begin{tabular}{lcl}
HTSC & cubic:linear in $B_{1g}$ & Reference\\
\tableline
Y-123 & 1 & Krantz {\it et al.}\cite{KraCar}\\
Y-123 & 1 & Hackl {\it et al.}\cite{Hacetal}\\
Y-123 & 0.35 & BZ integration\\
Bi-2212 & 0.07 & Staufer {\it et al.}\cite{Staetal}\\
\end{tabular}

\end{table}

\begin{table} 

\caption[(caption tab. 2)]{%
Comparison of the experimental peak scattering efficiencies 
given in units
of $10^{-8}\,{\rm cm}\,{\rm cm}^{-1}\,{\rm sr}^{-1}$ to the
theoretical predictions (from Fermi surface integrations,
Ref.~\protect\onlinecite{CarStr}, as well as Brillouin zone
integrations, present work) for Y-123 and Y-124.}
\label{peakh}

\begin{tabular}{crrrrrr}
{\bf Y-123} & 
\multicolumn{2}{c}{FS integration\cite{CarStr}} & 
\multicolumn{2}{c}{BZ integration} & 
\multicolumn{2}{c}{Experiment\cite{KraCar}} \\
Polarization & 
absolute & relative & 
absolute & relative & 
absolute & relative \\
\tableline
$yy$     &20.0&1.00 &19.6&1.00& 40&1.00 \\
$xx$     &28.0&1.40 & 7.2&0.37& 19&0.48 \\
$xy$     & 3.0&0.15 & 2.5&0.13&  4&0.10 \\
$x'x'$   &    &     & 5.0&0.26& 26&0.65 \\
$x'y'$   & 4.8&0.24 &10.6&0.54& 12&0.30 \\
$A_{1g}$ &19.2&0.96 & 3.0&0.15& 18&0.45 \\ 
\end{tabular}

\begin{tabular}{crrrrrr}
{\bf Y-124} & \multicolumn{2}{c}{FS integration\cite{CarStr}} & 
\multicolumn{2}{c}{BZ integration} & 
\multicolumn{2}{c}{Experiment\cite{DonKir}} \\
Polarization & 
absolute & relative & 
absolute & relative & 
absolute & relative \\
\tableline
$yy$     & 6.3&1.00 & 4.4&1.00&18.0&1.00 \\
$xx$     & 1.5&0.24 & 1.4&0.32& 7.2&0.40 \\
$xy$     & 1.1&0.17 & 0.5&0.11& 2.6&0.14 \\
$x'x'$   &    &     & 1.4&0.32&12.0&0.66 \\
$x'y'$   & 2.8&0.44 & 2.3&0.52& 5.6&0.31 \\
$A_{1g}$ & 1.1&0.17 & 1.0&0.23& 6.9&0.38 \\ 
\end{tabular}

\end{table}


\begin{figure}
\caption[(caption fig. 1)]{%
Incorporation of screening effects into the theory of Raman scattering
by electronic excitations in HTSC. The grey shaded bubbles are sums of
ladders contracted with vertices~$a$ and~$b$. Wavy lines correspond to
the long-range Coulomb interaction, dashed lines to the attractive
pairing interaction.  The equation on the last line corresponds to
Eq.~(\protect\ref{chir}).}
\label{screening}
\end{figure}

\begin{figure}
\caption[(caption fig. 2)]{%
Results from the BZ integration for Y-123. Given in the five panels
are absolute efficiencies for electronic Raman scattering. The upper
three curves are labelled using the irreducible representation
($D_{4h}$) of the scattering mass, the lower two panels with the
polarization geometry. Each of the five panels contains the total
absolute Raman efficiency according to Eq.~(\protect\ref{rameff})
and~(\protect\ref{fdt}) and its two constituents, the unscreened and
the screening part according to Eq.~(\protect\ref{chir}).}
\label{Y123}
\end{figure}

\begin{figure}
\caption[(caption fig. 3)]{%
Results from the Brillouin zone integration for Y-124. For details see
the caption of Fig.~\ref{Y123}.}
\label{Y124}
\end{figure}

\begin{figure}
\caption[(caption fig. 4)]{%
Experimental Raman scattering efficiencies for Y-123 from
Ref.~\protect\onlinecite{KraCar}.  The vertical scales are absolute
Raman efficiencies, measured at $T=10\,{\rm K}$ and an exciting laser
wavelength of $\lambda=488\,{\rm nm}$ (Note that a scale error found
in Ref.~\onlinecite{KraCar} has been corrected). The $A_{1g}$
component extracted according to
$I_{A_{1g}}=(I_{xx}+I_{yy})/2-I_{x'y'}$ is plotted in the lower panel
together with the
quasitetragonal $B_{1g}$ and $B_{2g}$ components.}
\label{Y123exp}
\end{figure}

\begin{figure}
\caption[(caption fig. 5)]{%
Upper panel: experimental absolute Raman efficiencies given for the
five specified polarization configurations for Y-124 from
Ref.~\protect\onlinecite{DonKir}. These data are taken at $T=10\,{\rm
K}$ with an exciting laser wavelength of $\lambda=514.5\,{\rm nm}$.
Lower panel: smoothed curves and the $A_{1g}$ spectrum additionally
extracted from the former. In both panels, consecutive offsets of
$0,1,\ldots,4\times 2.5 \,{\rm cm^{-1}sr^{-1}cm}$ were used.}
\label{Y124exp}
\end{figure}


\def\centerline#1{\hbox to\textwidth{\hfil#1\hfil}}

\newpage

\vbox{%
\epsfxsize0.9\textwidth
\epsfbox{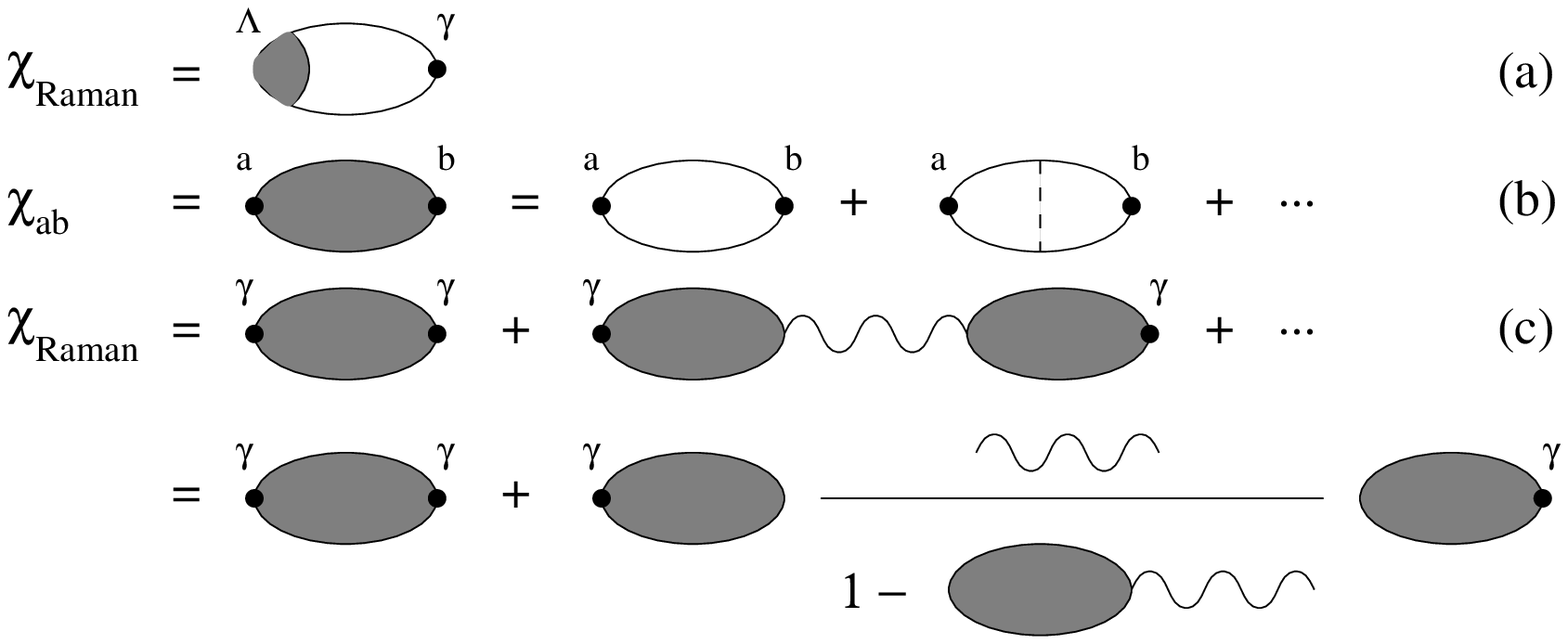}
\centerline{Figure~1}}
\newpage

\vbox{%
\epsfxsize0.85\textwidth
\epsfbox{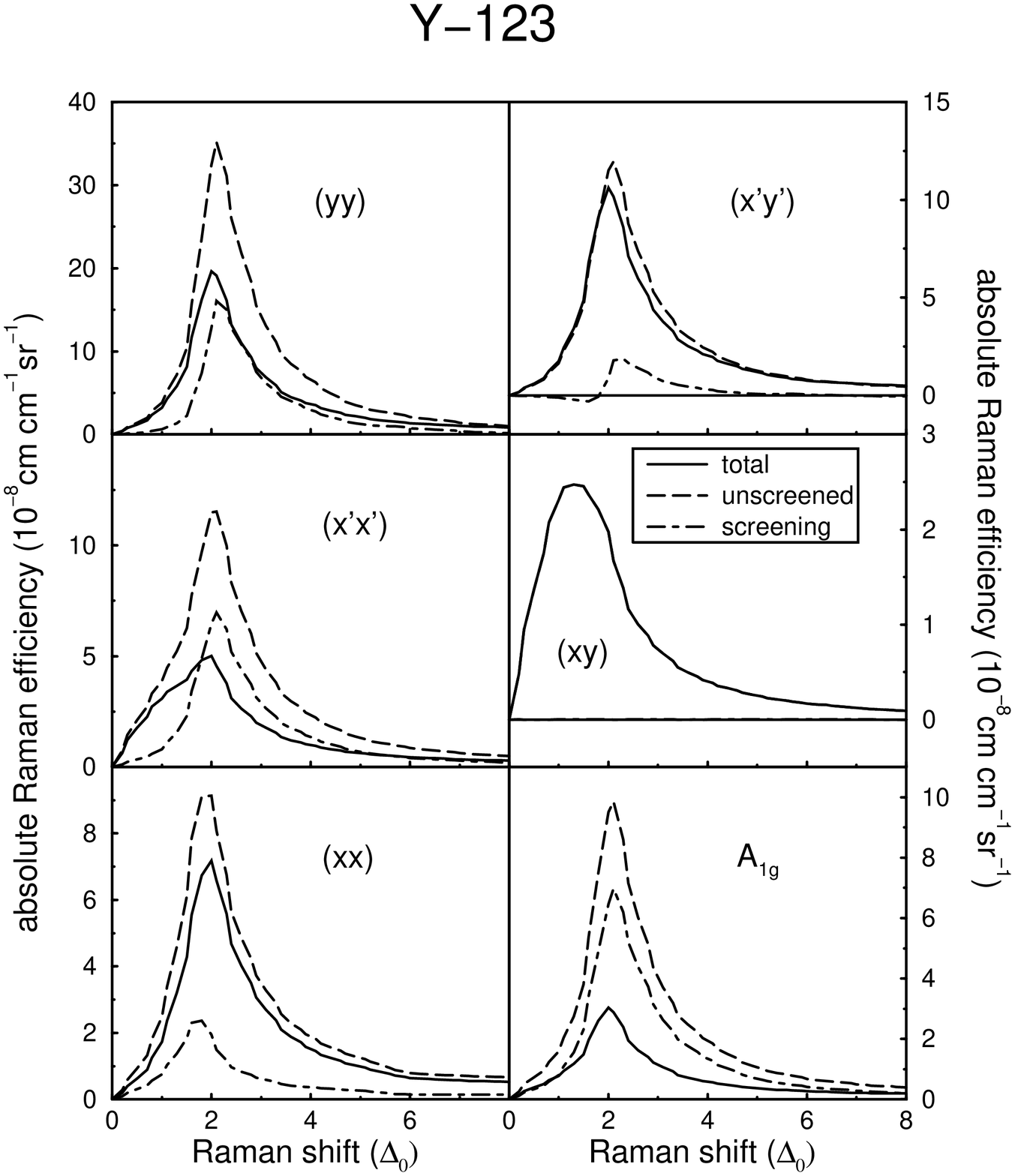}
\centerline{Figure~2}}
\newpage

\vbox{%
\epsfxsize0.85\textwidth
\epsfbox{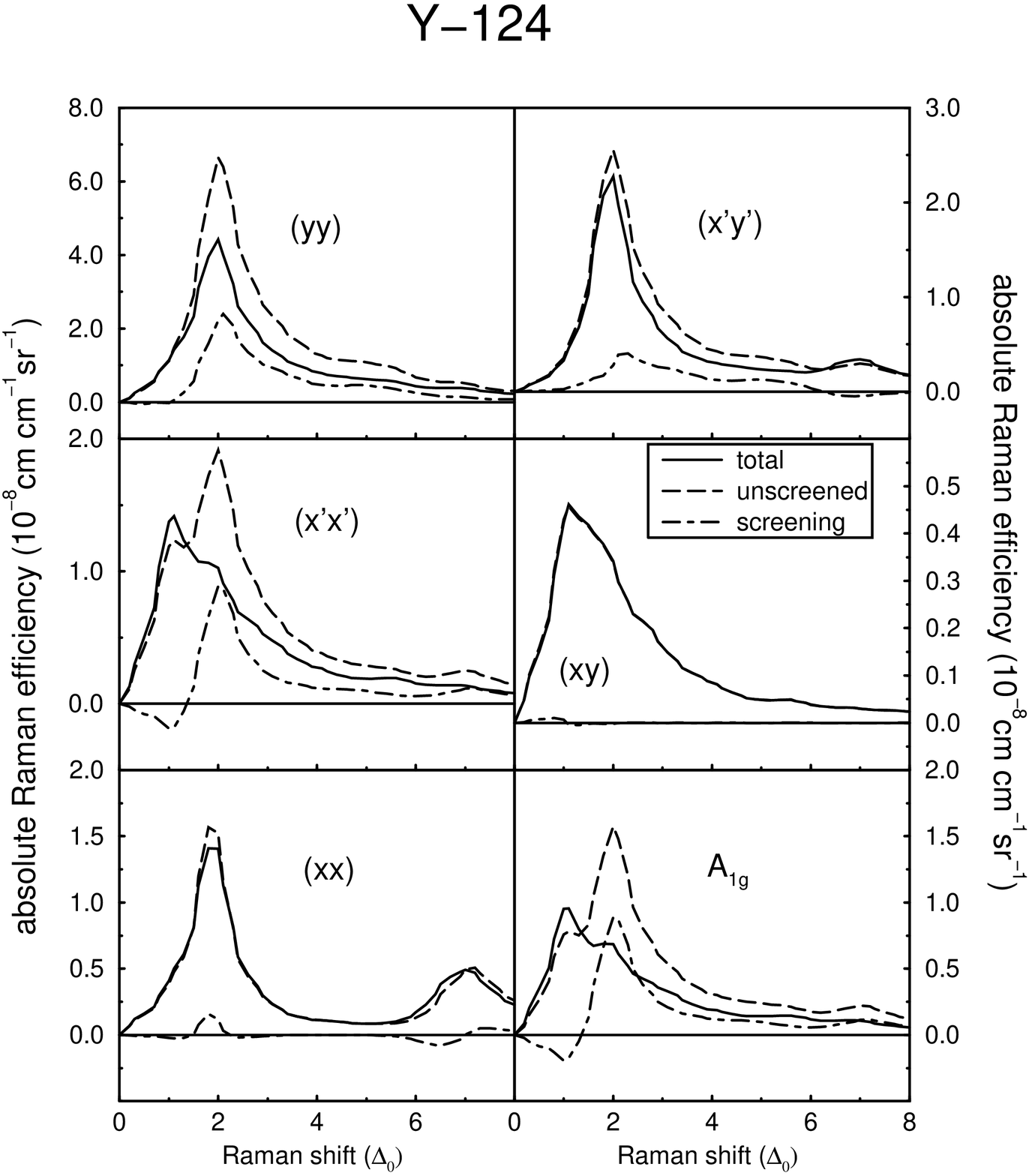}
\centerline{Figure~3}}
\newpage

\vbox{%
\epsfxsize0.7\textwidth
\centerline{\epsfbox{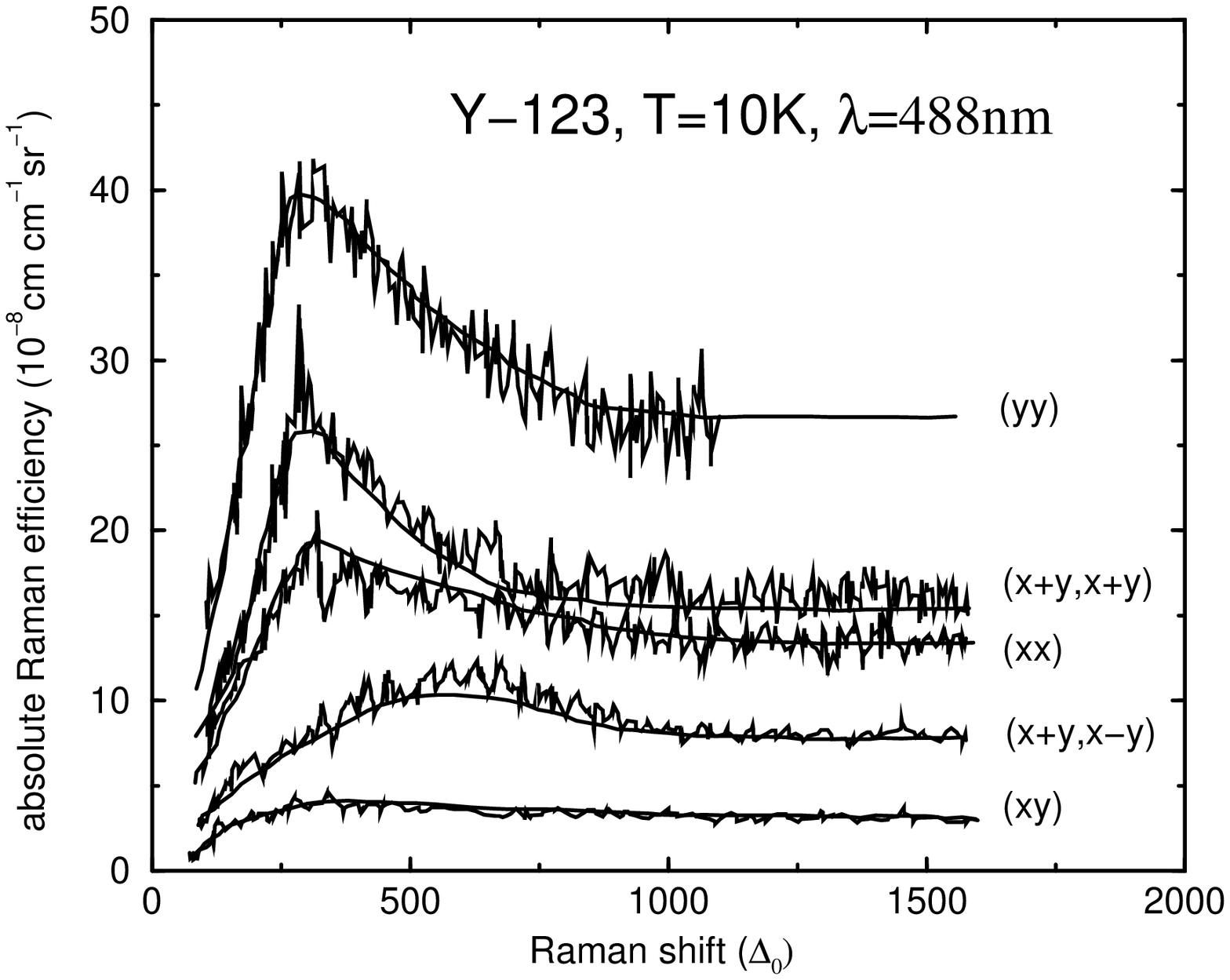}}
\epsfxsize0.7\textwidth
\centerline{\epsfbox{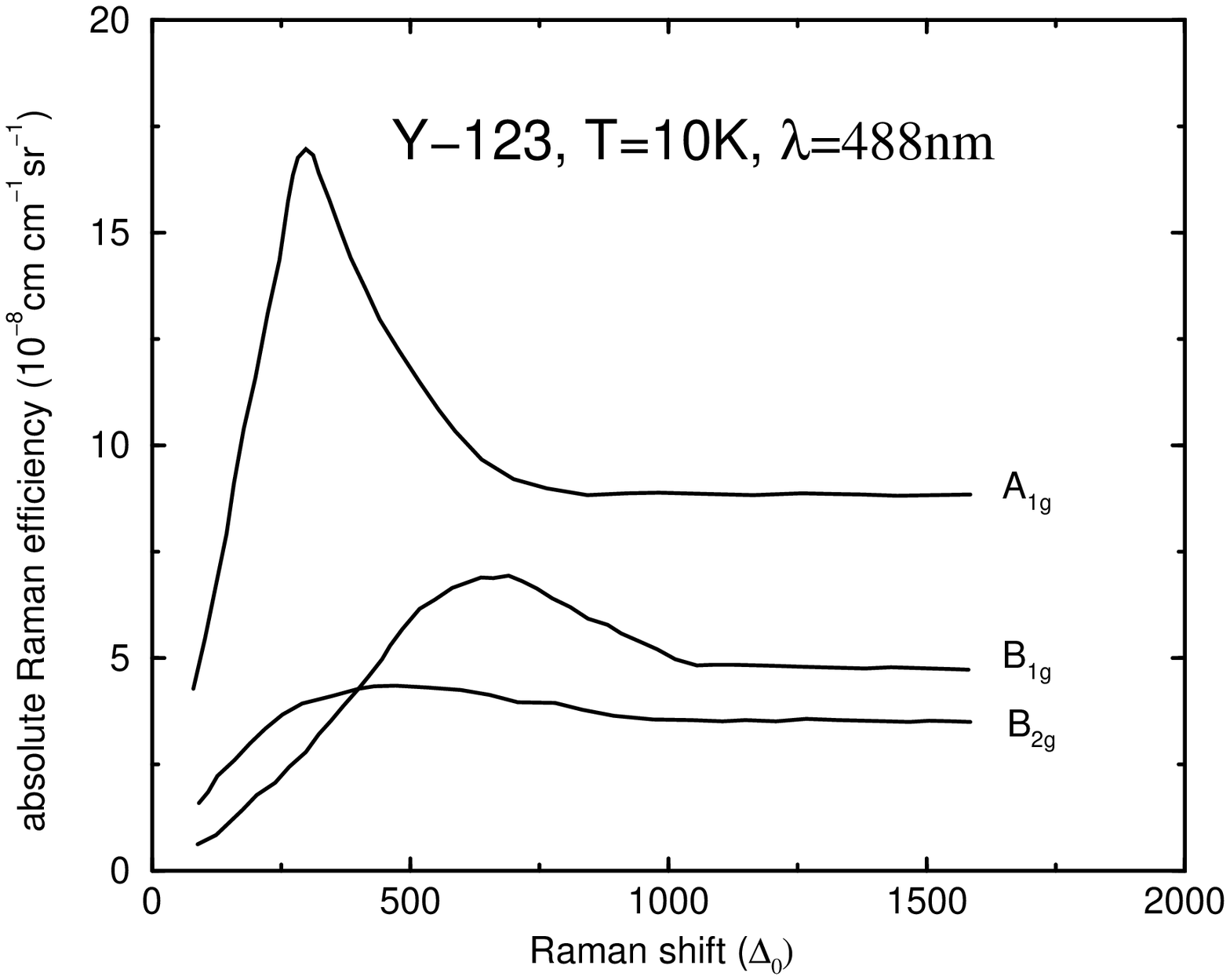}}
\centerline{Figure~4}}
\newpage

\vbox{%
\epsfxsize0.7\textwidth
\centerline{\epsfbox{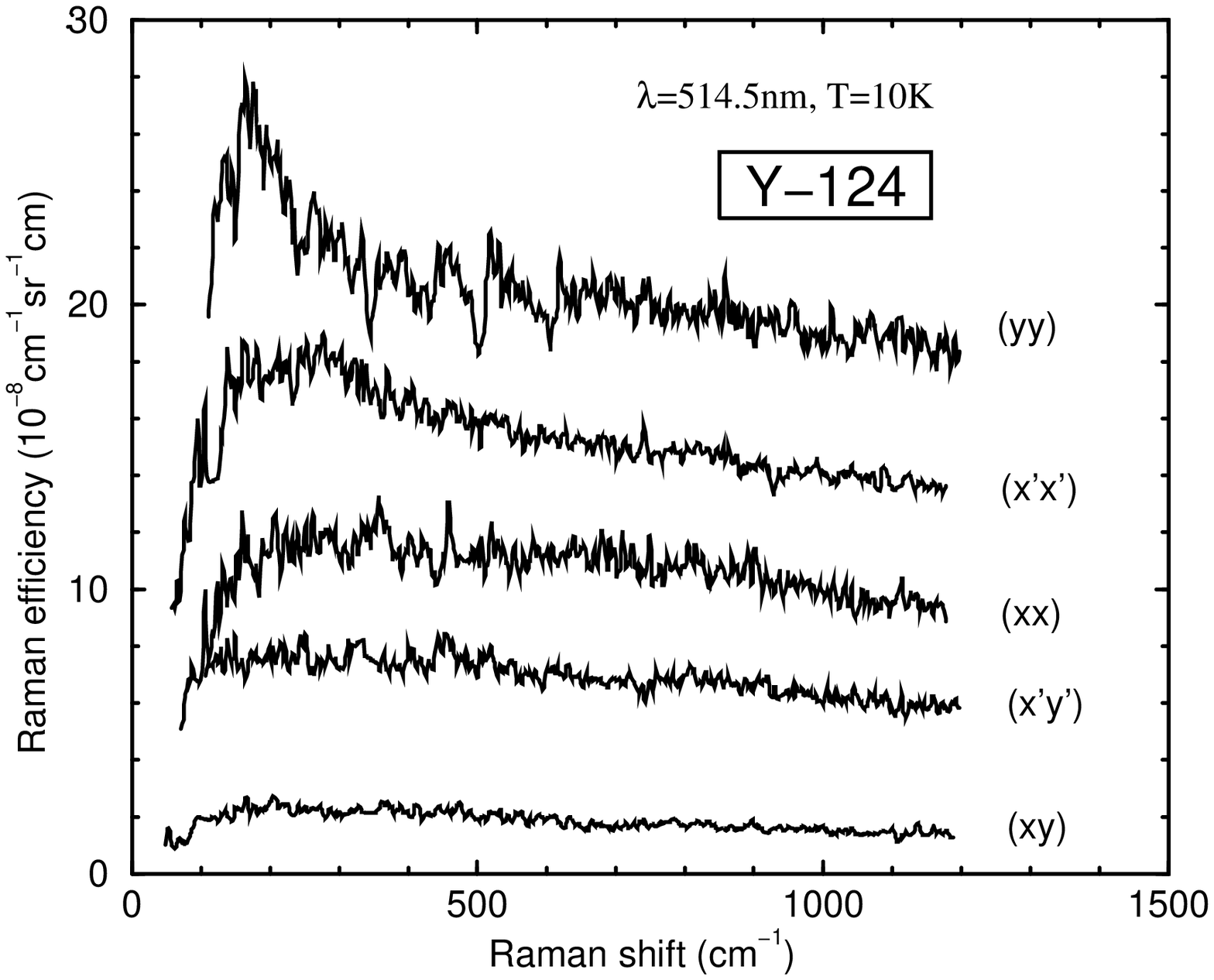}}
\epsfxsize0.7\textwidth
\centerline{\epsfbox{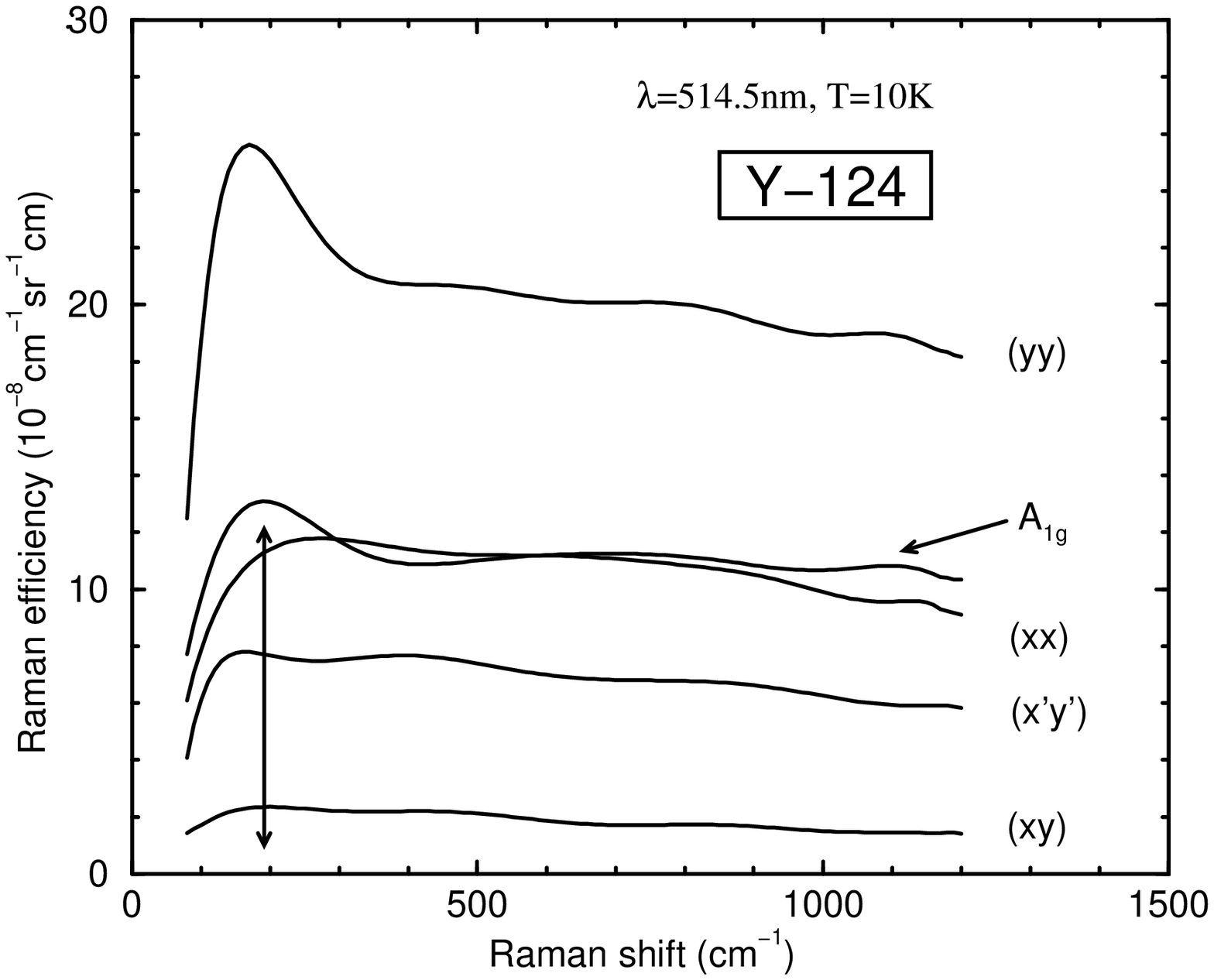}}
\centerline{Figure~5}}

\end{document}